\documentclass[12pt]{article}%
\pdfoutput=1
\usepackage[nosort]{cite}
\usepackage{graphicx}
\usepackage{multicol}
\usepackage{amsfonts}
\usepackage{amsthm}
\usepackage{amssymb}
\usepackage{amsmath}
\usepackage{singleauthor}
\usepackage{afterpage}
\usepackage{setspace}
\usepackage{verbatim}
\usepackage{slashed}
\usepackage{color}
\usepackage{longtable}
\usepackage{float}
\usepackage{subcaption}
\usepackage{epsfig}
\usepackage{bm}
\usepackage{enumerate}
\usepackage{epstopdf}
\usepackage[enableskew, vcentermath]{youngtab}
\usepackage{adjustbox}
\newsavebox{\mysavebox}

\usepackage[margin=1in]{geometry}
\usepackage{titletoc}%
\usepackage{hyperref}

\newtheoremstyle{named}{}{}{\itshape}{}{\bfseries}{.}{.5em}{#3}
\theoremstyle{named}
\newtheorem*{namedconjecture}{Conjecture}

\newcommand{\ba}{\begin{eqnarray}}
\newcommand{\ea}{\end{eqnarray}}

\newcommand{\cM}{\mathcal M}

\newcommand{\mKK}{m_{\textrm{KK}}}
\newcommand{\Mstring}{M_{\textrm{string}}}
\newcommand{\lH}{\lambda_{H}}
\newcommand{\lstring}{\lambda_{\textrm{string}}}
\newcommand{\lam}{\lambda}

\newcommand{\lKK}{\lambda_{\textrm{KK}}}
\newcommand{\lm}{\lambda_{\textrm{lightest}}}

\newcommand{\e}{\mathrm{e}}
\newcommand{\rmd}{\mathrm{d}}

\def\be{\begin{equation}}
\def\ee{\end{equation}}

\makeatletter \@addtoreset{equation}{section} \makeatother

\begin{document}

\date{\today}

\title{Asymptotic Scalar Field Cosmology in \\
String Theory}

\institution{BERKELEY}{Physics Department, University of California, Berkeley CA 94720 USA}

\authors{Tom Rudelius\worksat{\BERKELEY}\footnote{e-mail: {\tt rudelius@berkeley.edu}}}

\abstract{
Asymptotic (late-time) cosmology depends on the asymptotic (infinite-distance) limits of scalar field space in string theory. Such limits feature an exponentially decaying potential $V \sim \exp(- c \phi)$ with corresponding Hubble scale $H \sim \sqrt{\dot \phi^2 + 2 V} \sim \exp(-  \lH \phi)$, and at least one tower of particles whose masses scale as $m \sim \exp( - \lam \phi)$, as required by the Distance Conjecture. In this paper, we provide evidence that these coefficients satisfy the inequalities $\sqrt{(d-1)/(d-2)} \geq \lH  \geq \lm \geq 1/\sqrt{d-2}$ in $d$ spacetime dimensions, where $\lm$ is the $\lambda$ coefficient of the lightest tower. This means that at late times, as the scalar field rolls to $\phi \rightarrow \infty$, the low-energy theory remains a $d$-dimensional FRW cosmology with decelerated expansion, the light towers of particles predicted by the Distance Conjecture remain at or above the Hubble scale, and both the strong energy condition and the dominant energy condition are satisfied.
}

\maketitle

\tableofcontents

\enlargethispage{\baselineskip}

\setcounter{tocdepth}{2}

\section{Introduction}\label{sec:INTRO}

Recent years have seen a resurgence of interest in scalar field cosmology within quantum gravity. A series of provocative papers have raised the possibility of bounds on first and second derivatives of scalar field potentials, which in turn have generated lively debates over the existence/nonexistence of metastable de Sitter vacua, inflation, and quintessence in quantum gravity. 

Yet despite the heroic efforts of string theorists to settle these debates, the existence or non-existence of metastable de Sitter vacua in string theory--as well as the existence of inflation and quintessence--remains ``a matter of conjecture'' \cite{Dine:2020vmr}. The heart of the issue traces back to the observation of Dine and Seiberg \cite{Dine:1985he}, who pointed out that de Sitter vacua in string theory necessarily lie outside the regime of parametric control, as multiple terms in a perturbative expansion must compete to produce such a minimum. Very little is known outside of such parametrically controlled regimes in string theory, and as a result, it is very difficult to make convincing claims one way or the other about de Sitter, inflation, and dark energy in string theory.

In light of these difficulties, many string theorists have refocused their efforts on a systematic characterization of weakly-coupled regimes, where perturbative control permits a much greater degree of rigor. Since couplings in string theory are controlled by vacuum expectation values (vevs) of scalar fields, these regimes are associated with limits in which certain (noncompact) scalar fields are taking to infinity. In such limits, string theory seems to admit a number of universal features, which have led to the formulation of a handful of ``swampland'' conjectures regarding these features.

One conjecture of this form is the Distance Conjecture \cite{Ooguri:2006in}, which says:
\vspace{.2cm}
           \begin{namedconjecture}[The Distance Conjecture]
Let $\cM$ be the scalar field moduli space of a quantum gravity theory in $d \geq 4$ dimensions, parametrized by vacuum expectation values of massless scalar fields. Compared to the theory at some point $p_0 \in \mathcal{M}$, the theory at a point $p \in \mathcal{M}$ has an infinite tower of particles, each with mass scaling as
\be
m \sim \exp( -\lambda ||p - p_0|| )\,,
\label{DCdef}
\ee 
where $||p - p_0||$ is the geodesic distance in $\mathcal{M}$ between $p$ and $p_0$, and $\lambda$ is some order-one number in Planck units $(8 \pi G = \kappa_d^2 = 1)$.
            \end{namedconjecture}
    \vspace{.1cm}
\noindent
In its most conservative formulation, the Distance Conjecture applies only to exactly massless scalar fields parametrizing the moduli space of a supersymmetric theory. However, many have argued that the conjecture should apply more generally to any infinite-distance limit in scalar field space \cite{Ooguri:2006in, Klaewer:2016kiy}, so that a tower of particles becomes light in exponential fashion whenever the vev of a noncompact scalar field is taken to infinity. In this paper, we will discuss both formulations of the conjecture, but we will reserve the term ``moduli'' for massless scalar fields and ``scalar field moduli space'' (or simply ``moduli space'') for the manifold parametrized by these massless scalar fields. In contrast, we will refer to the space parametrized by vevs of (possibly massive) scalar fields as simply ``scalar field space.''

Exponential behavior seems to be a universal feature of scalar field potentials in asymptotic limits of scalar field space as well. This was pointed out long ago by Dine and Seiberg \cite{Dine:1985he}, but more recently it has been codified in the asymptotic de Sitter Conjecture \cite{Obied:2018sgi, Hebecker:2018vxz}, which we define as follows:
\vspace{.2cm}
           \begin{namedconjecture}[The Asymptotic de Sitter Conjecture]
Given any point in scalar field space $p_0 \in \mathcal{M}$, there exists some large radius $R$ such that for $||p-p_0|| > R$ (i.e., in an asymptotic regime of scalar field space), the scalar field potential at $p$ satisfies
\begin{equation}
\frac{|\nabla V|}{V} \geq c_{\textrm{min}}\,,
\end{equation}
provided $V(p) > 0$.
            \end{namedconjecture}
    \vspace{.1cm}
\noindent

A bound of the form $|\nabla V|/V \geq c_{\textrm{min}}$ leads to an exponentially decaying potential of the form $V \sim \exp( - c \phi )$ in the limit $\phi \rightarrow \infty$, with $c \geq c_{\textrm{min}}$.

Finally, one other conjecture that will play an important role in our discussion is the Emergent String Conjecture \cite{Lee:2019wij,Lee:2019xtm}. This conjecture holds the following: 
\vspace{.2cm}
           \begin{namedconjecture}[The Emergent String Conjecture]
Any infinite-distance limit in scalar field moduli space is either a decompactification limit or an emergent string limit, in which the tension of a fundamental string asymptotically approaches zero.
            \end{namedconjecture}
    \vspace{.1cm}
\noindent
Like the Distance Conjecture, the Emergent String Conjecture in its most conservative formulation applies only to moduli spaces of exactly massless scalar fields. Once again, however, it is plausible that it applies more generally to infinite-distance limits of scalar field space \cite{Klaewer:2020lfg}, as we discuss further in \S\ref{sec:KKRED} below. 

Each of these conjectures is supported by a wide array of examples in string theory and bottom-up arguments from supergravity and/or effective field theory (see e.g. \cite{Cicoli:2017axo, Blumenhagen:2018nts, Grimm:2018ohb, Corvilain:2018lgw, Heidenreich:2018kpg, Erkinger:2019umg, Joshi:2019nzi, Rudelius:2021oaz, EnriquezRojo:2020hzi, Rudelius:2021azq, Etheredge:2022opl}). When they are combined, a simple, coherent picture of infinite-distance limits in scalar field space begins to emerge. 
%
%
First off, the asymptotic de Sitter Conjecture suggests that the potential in asymptotic limits of scalar field space should, at leading order, take the form $
V \sim \exp( - c_i \phi^i )$,
where $\phi^i$ are scalar fields and $c_i$ are numerical coefficients. By appropriate field redefinitions, this may be brought to the form
\begin{equation}
V \sim \exp( - c   \hat \phi )\,,
\label{Vexp}
\end{equation}
where $\hat \phi$ is a canonically-normalized scalar field measuring the proper field distance traversed, and more generally we use a hat $\hat{\cdot}$ throughout this paper to indicate a canonically normalized scalar field.

Consider now a homogenous scalar field ($\hat \phi = \hat \phi(t)$) rolling towards $\hat \phi = \infty$ in an asymptotic limit of scalar field space. As we show in Appendix \ref{App}, FRW cosmology then implies a Hubble scale $H$ of the form
\begin{equation}
H \sim \sqrt{\dot{ \hat \phi}^2 + 2V}  \sim  \exp( - \lambda_H   \hat \phi ) \,,~~~~\lambda_H = \min \left( \frac{c}{2}, \sqrt{\frac{d-1}{d-2}} \right)\,,
\label{HV}
\end{equation}
where $\lH = \sqrt{(d-1)/(d-2)}$ whenever the kinetic energy of $\hat \phi$ dominates its potential energy, which indeed occurs whenever $c \geq 2 \sqrt{(d-1)/(d-2)}$.

Then, the Distance Conjecture implies the existence of a tower of particles whose masses scale exponentially in the limit $\hat \phi \rightarrow \infty$ as $m \sim \exp ( - \lam \hat \phi )$. In typical cases, there will in fact be multiple towers of this form, each with their own scaling coefficient $\lambda_i$, i.e., $m_i \sim \exp (-\lam_i \hat \phi)$. In this work, we will be most interested in the lightest tower, or equivalently the tower with the largest $\lambda_i$, so we define
\begin{equation}\lm = \max_i (\lambda_i) \,.
\end{equation}
The Emergent String Conjecture further implies that this lightest tower is either a tower of Kaluza-Klein modes or a tower of string oscillator modes.


Putting this together, we expect that the late-time behavior of a scalar field rolling in a asymptotic regime in scalar field space in string theory will be characterized by two energy scales, $m$ and $H$, each of which scale exponentially with proper field distance in Planck units:
\begin{align}
m \sim \exp(  - \lm  \hat \phi ) \,,~~~~  H \sim \exp( - \lH   \hat \phi )\,.
\end{align}

The associated late-time cosmological evolution depends crucially on the question of which of these energy scales is smallest, or equivalently, on the question of which of the $\lambda$ coefficients is largest. If $\lm > \lH$, then the late-time cosmology will depend on the nature of the lightest tower of particles, which by the Emergent String Conjecture ought to be either a Kaluza-Klein tower or a tower of string oscillator modes. If this tower is a Kaluza-Klein tower, then the Kaluza-Klein scale $\mKK$ will drop below the the Hubble scale, so the size of the compactified dimensions $\mKK^{-1}$ will be larger than the horizon size $H^{-1}$, and the cosmology can no longer be viewed as a $d$-dimensional spacetime. If the lightest tower is a tower of string oscillator modes, then the string scale will eventually drop below the Hubble scale, and effective field theory will break down altogether as the universe enters a stringy phase. If $\lH > \lm$, then the universe will continue to expand indefinitely, remaining a $d$-dimensional cosmology since the Kaluza-Klein scale is above the Hubble scale. If additionally $\lH > 1/\sqrt{d-2}$, then the universe will undergo decelerated expansion $\ddot a < 0$, whereas if $\lH < 1/\sqrt{d-2}$ the universe will undergo accelerated expansion $\ddot a > 0$.

In this paper, we argue that in $d$ spacetime dimensions, the $\lambda$ coefficients satisfy the relationships
\begin{align}
\sqrt{\frac{d-1}{d-2}} \geq \lH \geq \lm  \geq \frac{1}{\sqrt{d-2}} \,.
\label{propbound}
\end{align}
If the latter two inequalities are strictly satisfied, i.e. $\lH > \lm > \frac{1}{\sqrt{d-2}}  $, the result is indefinite decelerated expansion of the universe.

Note that in theories with multiple scalar fields, the proposed bounds \eqref{propbound} do not apply to every asymptotic direction in scalar field space. The field $\hat \phi$ was defined above to be the proper distance along a path traversed by a rolling scalar field, which in asymptotic regimes of scalar field space is dominated by an exponential as in \eqref{Vexp}. In other directions in scalar field space, the directional derivative of the potential may be arbitrarily small or even positive, but at sufficiently late times and sufficiently far out in scalar field space, the field will not roll in these directions, so they are irrelevant from the perspective of the late-time cosmology.

The remainder of this paper is structured as follows: in \S\ref{sec:KKRED}, provide evidence for our proposed bounds in the context of Kaluza-Klein reduction. In \S\ref{sec:ST}, we provide evidence from string theory. In \S\ref{sec:COSMO}, we study the implications of these bounds for cosmology. In \S\ref{sec:CONC}, we conclude with a discussion of possible directions for further research.


\section{Kaluza-Klein Reduction}\label{sec:KKRED}

In this section, we provide evidence for our proposed bounds $\lH \geq \lm \geq 1/\sqrt{d-2}$ in the context of Kaluza-Klein reduction. This section is organized as follows: in \S\ref{SDC}, we review and extend the evidence for the sharpened Distance Conjecture, which implies $\lm \geq 1/\sqrt{d-2}$ in any direction in scalar field space. In \S\ref{dSC}, we review the Kaluza-Klein reduction argument for the asymptotic de Sitter Conjecture, which implies $\lH \geq 1/\sqrt{d-2}$. In \S\ref{HKK}, we present novel arguments in the context of Kaluza-Klein reduction for the proposed bound $\lH \geq \lm$.

\subsection{A Sharpened Distance Conjecture}\label{SDC}

In this subsection, we review the analysis of \cite{Etheredge:2022opl}, which proposed the bound $\lambda \geq 1/\sqrt{d-2}$ for the coefficient of the lightest tower in any infinite-distance limit of scalar field moduli space.

To begin, we consider a $D$-dimensional theory with a single, canonically normalized scalar field $\hat \phi$. We further suppose that in the limit $\hat \phi \rightarrow \infty$, there exists a tower of particles whose masses scale as
\be
m_{\text{part}}^{(D)}   \sim \exp ( - \kappa_D \lambda_D   \hat \phi  )\,.
\ee 
We then consider a Kaluza-Klein reduction ansatz to $d = D-n$ dimensions of the form
\begin{equation}
ds^2 = \e^{- \frac{2n}{d-2} \rho(x)} g_{\mu\nu}^{(d)} \rmd x^\mu \rmd x^\nu + \e^{\rho(x)} h_{lm}^{(n)} \rmd y^l \rmd y^m,
\label{ansatz}
\end{equation}
where $\rho(x)$ is the radion field, $\mu, \nu = 0,...,d-1$, and $l,m=1,...,n$. In the $d$-dimensional theory, there are then two towers of charged particles of interest for us. One is associated with the reduction of the $D$-dimensional tower, with masses that scale as
\begin{align}
m_{\textrm{part}}^{(d)}& \sim \exp \left( - \lambda_D \hat \phi -  \kappa_d   \sqrt{\frac{ n}{{(n+d-2)(d-2)} } }\hat{\rho}  \right) \,.  
\end{align}
The other consists of the Kaluza-Klein modes, with masses that scale as
\begin{align}
m_{\textrm{KK}}^{(d)}& \sim \exp \left(-  \kappa_d   \sqrt{\frac{ {n+d-2} }{{n(d-2)} } }\hat{\rho}  \right) \,.\label{nKK} 
\end{align}

\begin{figure}
\begin{center}
\center
\includegraphics[width=60mm]{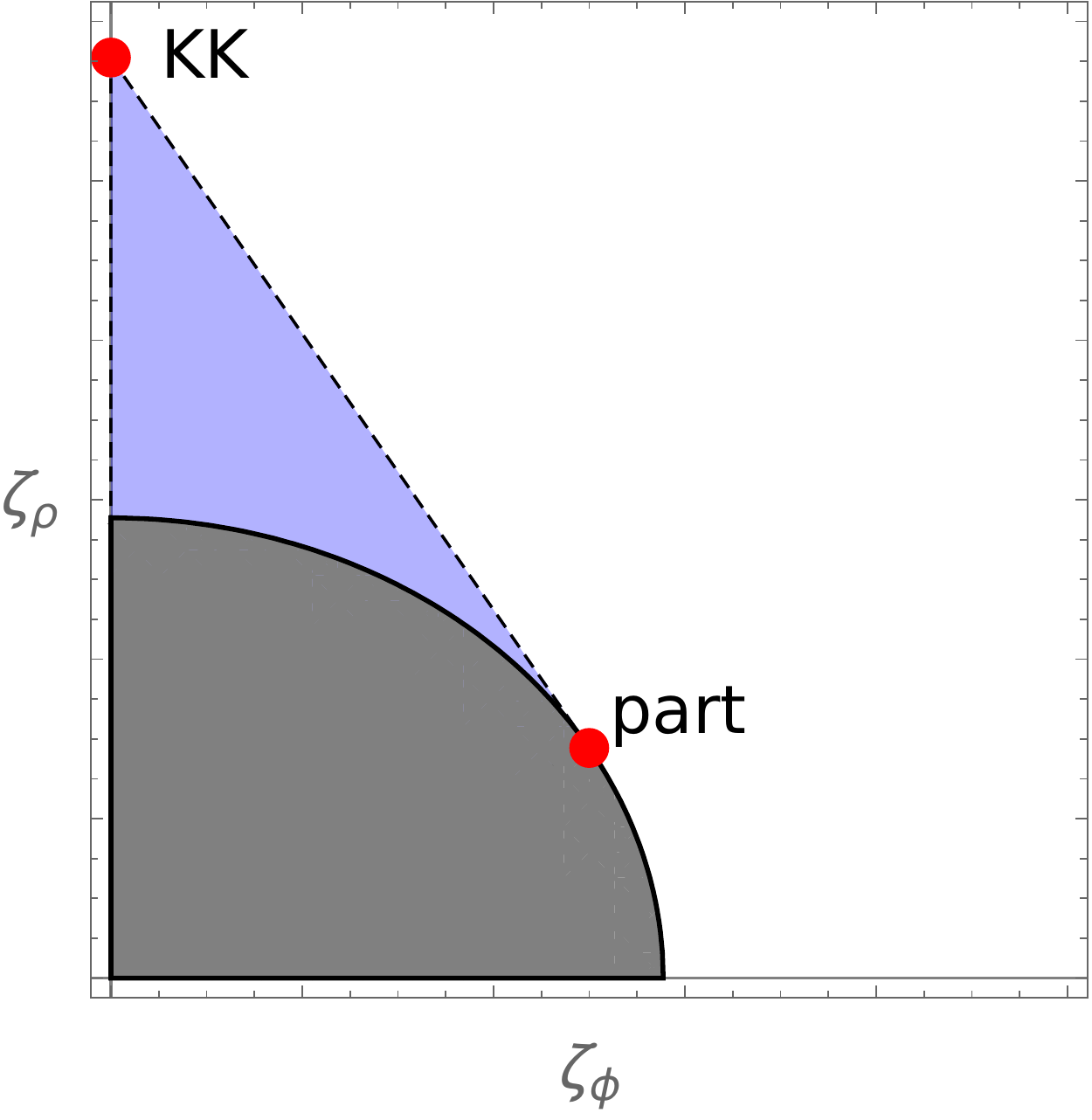}
\caption{Kaluza-Klein reduction and the convex hull condition. The convex hull generated by the vector $\vec\zeta_\text{KK}=(0,\sqrt{\frac{n+d-2}{n(d-2)}})$  $\vec\zeta_\text{part}=(\sqrt{\frac{1}{n+d-2}}, \sqrt{\frac{n}{(n+d-2)(d-2)}}  )$ lies outside the ball of radius $1/\sqrt{d-2}$, so the Distance Conjecture will be satisfied with a coefficient $\lm \geq 1/\sqrt{d-2}$ in every direction $\hat \rho, \hat \phi \rightarrow \infty$ with $\hat \rho/ \hat \phi \geq \sqrt{n/(d-2)}$. Figure adapted from \cite{Etheredge:2022opl}.}
\label{KKCHC}
\end{center}
\end{figure}

It is helpful to define the scalar charge-to-mass vector of a particle as
\be
\zeta_i\equiv - \frac 1{\kappa_d}\frac\partial{\partial \phi^i}\log m\,,
\label{zetavec}
\ee
Thus, in the case at hand, we have $\vec\zeta_\text{part}^{(D)} = (\lambda_D)$, and
\be
\vec\zeta_\text{part}^{(d)} = \begin{pmatrix}\lambda_D \\ \left( \frac{n}{(n+d - 2) (d - 2)} \right)^{1/2} \end{pmatrix} \,, \qquad \vec{\zeta}_{\textrm{KK}}^{(d)} =  \begin{pmatrix}0\\  \left( \frac{n+d-2}{n(d -2)}\right)^{1/2} \end{pmatrix}\,.
\label{ndimred}
\ee
Here, we notice that for $\lambda_D = 1/\sqrt{D-2}$, the scalar charge-to-mass vectors for the particles satisfy
\be
| \vec\zeta_\text{part}^{(D)}| = \frac{1}{\sqrt{D-2}} \,,~~~~ | \vec\zeta_\text{part}^{(d)}| = \frac{1}{\sqrt{d-2}} \,.
\ee
Furthermore, as shown in Figure \ref{KKCHC}, the convex hull generated by $\vec\zeta_\text{part}^{(d)}$ and $\vec{\zeta}_{\textrm{KK}}^{(d)}$ is actually tangent to the ball of radius $\lambda_d=1/\sqrt{d-2}$ for $\lambda_D = 1/\sqrt{D-2}$. This means that any limit with $\phi, \rho \rightarrow \infty$, $\rho / \phi \geq \sqrt{n/(n+d-2)}$ will feature a tower of light particles satisfying the Distance Conjecture with $\lambda \geq 1/\sqrt{d-2}$. We conclude that the value $\lambda_d = 1/\sqrt{d-2}$ is distinguished in that it is exactly preserved under dimensional reduction. This is a first hint towards the bound
\begin{equation}
\lm \geq \frac{1}{ \sqrt{ d-2 }} \,
\label{towerbound}
\end{equation}
on the exponential decay coefficient of the lightest tower in any infinite-distance limit in scalar field space. This was first pointed out in \cite{Etheredge:2022opl}, which gave several lines of evidence in favor of this bound.

\subsubsection{A toy model of dimensional reduction}\label{toy}

Of course, the scenario we have just considered is incomplete, since it does not include a tower of light particles in a limit with $\hat \rho \rightarrow - \infty$ or $\hat \phi \rightarrow - \infty$. In string theory, the presence of these towers is ensured by dualities. As a prototypical example, consider a $D$-dimensional theory with a fundamental string whose tension scales as $T_{\textrm{string}}^{+} \sim \exp(- 2 \hat \phi / \sqrt{D-2})$ in the limit $\hat \phi \rightarrow \infty$, and suppose that the limit $\hat \phi \rightarrow - \infty$ features a dual string with $T_{\textrm{string}}^{-} \sim \exp(2 \hat \phi / \sqrt{D-2})$. This is precisely the scaling behavior seen, for instance, in Type IIB string theory compactified on $T^n$ to $D=10-n$ dimensions, where $\hat \phi$ is the $D$-dimensional dilaton and $T_{\textrm{string}}^{\pm}$ represents the tension of the F/D-string, respectively.

After reduction to $d=D-1$ dimensions, we find several distinct towers of charged particles, five of which are especially interesting for our purposes. The first are Kaluza-Klein modes, which scale with the radion $\rho$ as in \eqref{nKK} with $n=1$. The second and third are string oscillator modes associated with the strings of tension $T_{\textrm{string}}^{+}$ and $T_{\textrm{string}}^{-}$. These descend from the string oscillator modes in the $D$-dimensional theory and have masses which scale as
\begin{equation}
\Mstring^\pm \sim \exp \left(  \mp \frac{\kappa_d}{\sqrt{D-2}}  \hat\phi -  \kappa_d   \frac{ 1}{\sqrt{(d-1)(d-2)}  }\hat{\rho}   \right) \,.
\label{Mstringscaling}
\end{equation}
Finally, there are two towers of winding modes, which come from wrapping the strings around the dimensional reduction circle. These have 
\begin{equation}
m_{\textrm{wind}}^\pm \sim \exp \left(  \mp  \kappa_d \frac{2 }{\sqrt{D-2}}  \hat\phi +  \kappa_d   \frac{ d-3}{\sqrt{(d-1)(d-2)}  }\hat{\rho}   \right) \,.
\end{equation}
In $d=4$ dimensions, there are also Kaluza-Klein monopoles, whose masses scale as
\begin{equation}
m_{\textrm{mon}}^\pm \sim \exp \left(  +  \kappa_d   \sqrt{\frac{3}{2}} \hat{\rho}   \right) \,,
\end{equation}
and in $d=5$ dimensions, there are oscillator modes of a Kaluza-Klein monopole string, whose masses scale as
\begin{equation}
m_{\textrm{mon. str.}}^\pm \sim \exp \left(  +  \kappa_d   \frac{1}{\sqrt{3}} \hat{\rho}   \right) \,.
\end{equation}
This leads to $\vec\zeta$-vectors of the form
\be
\vec \zeta_\text{KK}^{(d)}=\begin{pmatrix}0\\ \sqrt{\frac{d - 1}{d -2}} \end{pmatrix},\qquad \zeta_\text{string}^{(d)}=\begin{pmatrix}\pm \frac 1{\sqrt{d-1}}\\\frac{1}{\sqrt{(d - 1) (d - 2)}} \end{pmatrix} \,, \qquad\vec \zeta_\text{wind}^{(d)}=\begin{pmatrix}\pm \frac 2{\sqrt{d-1}}\\-\frac{d-3}{\sqrt{(d-1)(d-2)}}\end{pmatrix},
\ee
as well as 
\be
\vec \zeta_\text{mon}^{(4)}=\begin{pmatrix}0\\-\sqrt{\frac 32}\end{pmatrix}\,.
\ee
in four dimensions, and 
\be
\vec \zeta_\text{mon. str.}^{(5)}=\begin{pmatrix}0\\-\frac{1}{\sqrt{3}}\end{pmatrix}\,.
\ee
in five dimensions.

\begin{figure}
\begin{center}
\begin{subfigure}{0.475\textwidth}
\center
\includegraphics[width=50mm]{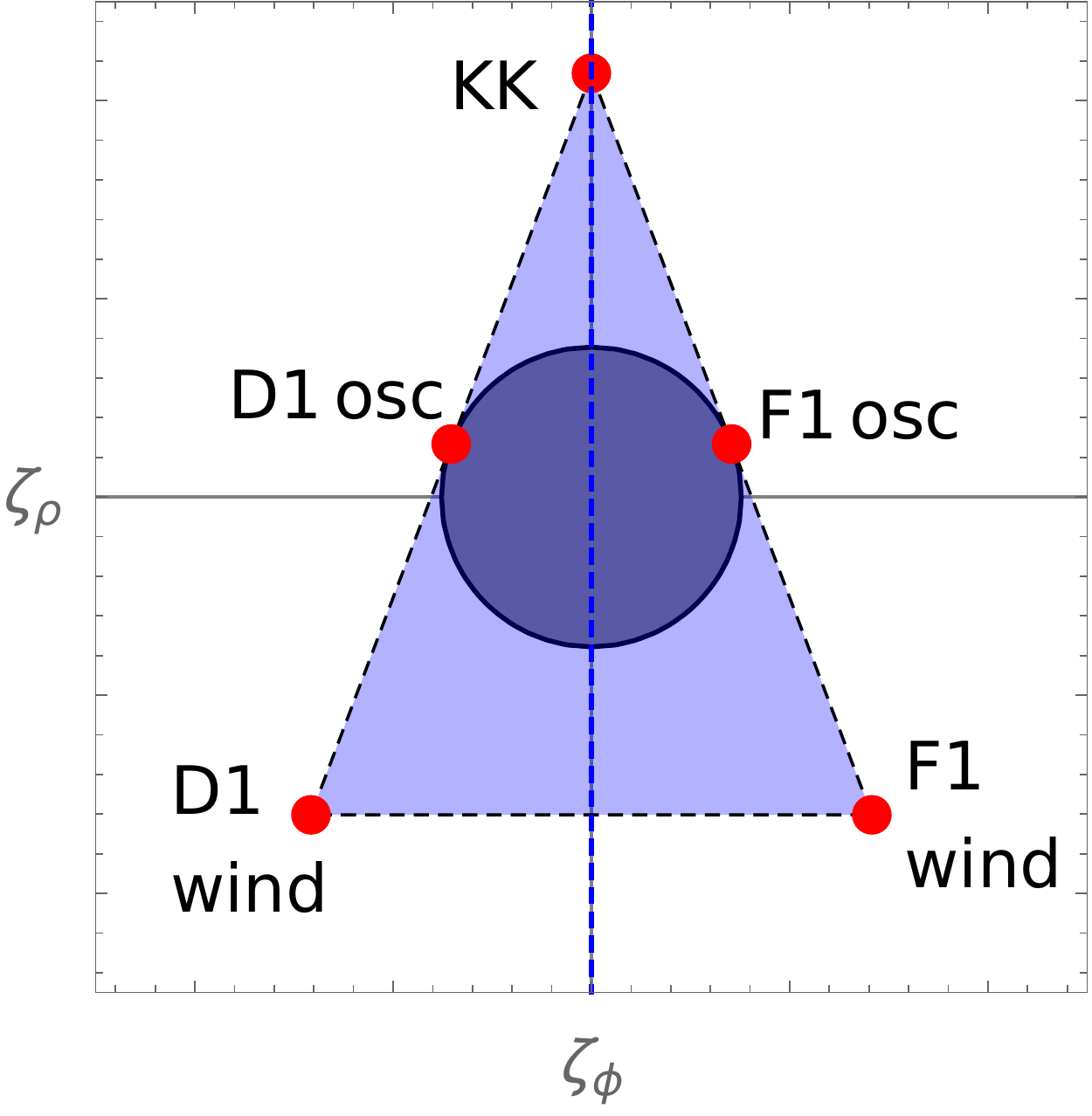}
\caption{$d>5$} \label{sfig:zetadgeq6}
\end{subfigure}
\begin{subfigure}{0.475\textwidth}
\center
\includegraphics[width=50mm]{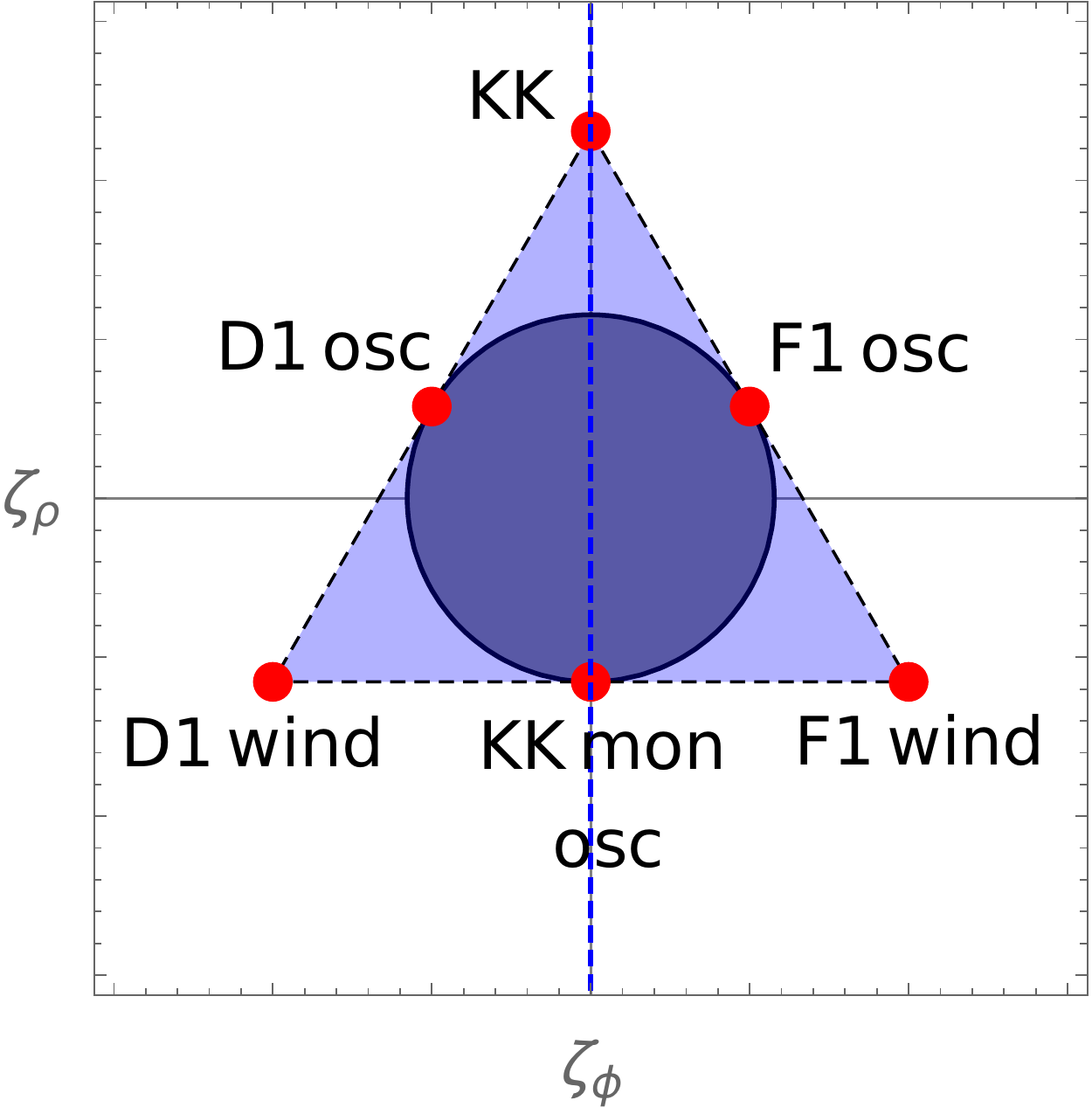}
\caption{$d=5$} \label{sfig:zeta5d}
\end{subfigure}
\hfill
\begin{subfigure}{0.475\textwidth}
\center
\includegraphics[width=50mm]{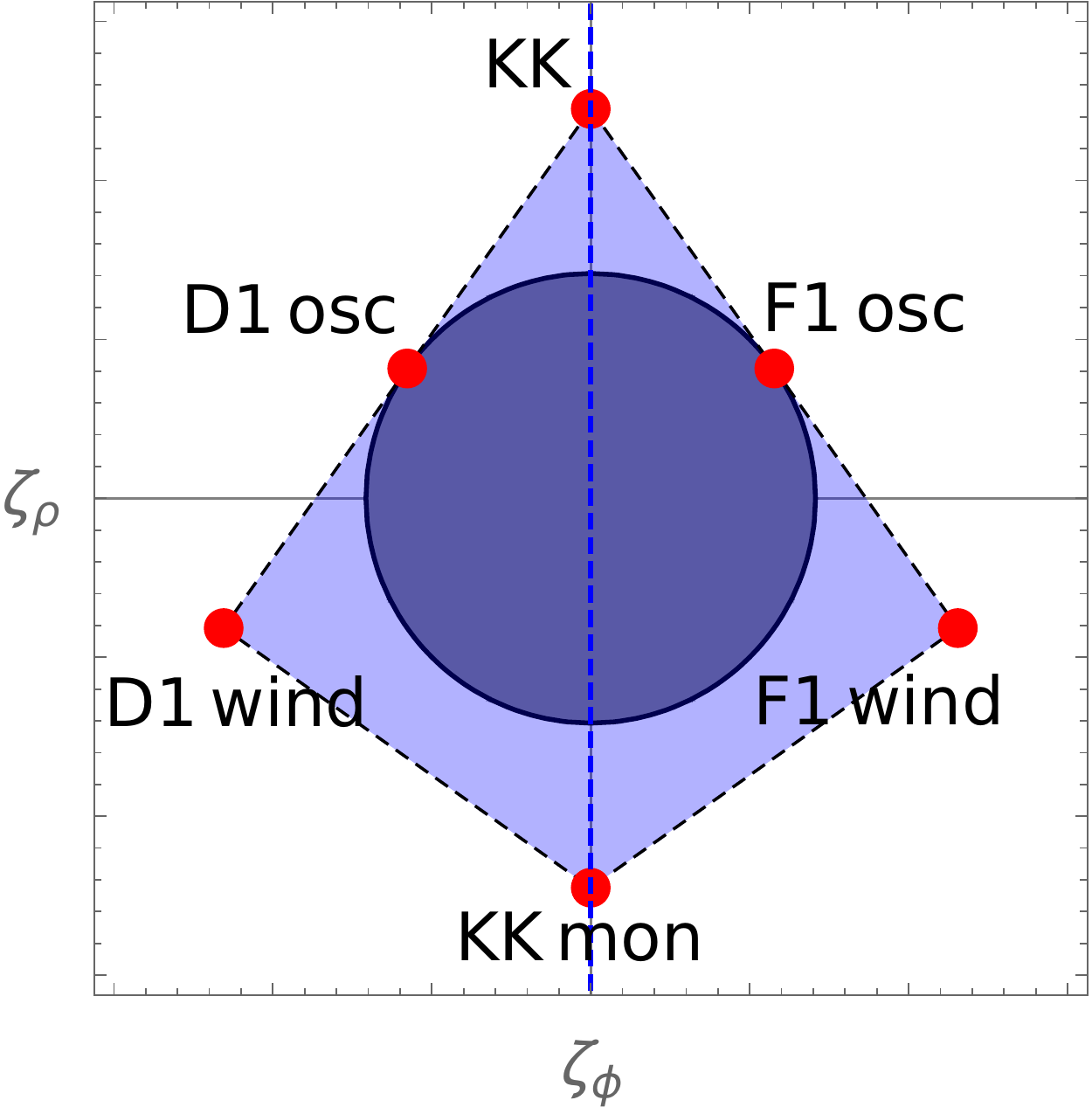}
\caption{$d=4$} \label{sfig:zeta4d}
\end{subfigure}
\caption{The convex hull condition in a toy model of dimensional reduction. The gray region in both figures is the ball of radius $1/\sqrt{d-2}$ centered at the origin. The convex hull generated by the $\vec{\zeta}$-vectors for the Kaluza-Klein modes, the winding string modes, and (in $d=4$ dimensions) the Kaluza-Klein monopole contains the ball of radius $1/\sqrt{d-2}$ (shown in gray). When the bound $\lm = 1/\sqrt{d-2}$ is saturated, so that the convex hull is tangent to the ball of radius $1/\sqrt{d-2}$, there is always a tower of string oscillator modes (which may come from the monopole string in $d=5$ dimensions). Figure adapted from \cite{Etheredge:2022opl}.}
 \label{FullKK}
\end{center}
\end{figure}

The convex hull of these $\vec\zeta$-vectors is shown in Figure \ref{FullKK}, along with the ball of radius $|\vec{\zeta}| = 1/\sqrt{d-2}$, which is completely contained in the convex hull. From this, we observe several things:
\begin{itemize}
\item The Distance Conjecture is satisfied in every direction in scalar field moduli space by at least one tower with $\lambda \geq 1/\sqrt{d-2}$. This follows from the fact that the convex hull of the $\vec\zeta$-vectors contains the ball of this radius--a criterion we refer to as the ``convex hull condition.'' 
\item The generators of the convex hull have length $|\vec \zeta| = \sqrt{(d-1)/(d-2)}$, which is the appropriate length for a tower of Kaluza-Klein modes for a decompactification to $D=d+1$ dimensions. This is to be expected from the Emergent String Conjecture \cite{Lee:2019wij,Lee:2019xtm, Alvarez-Garcia:2021pxo}, which holds that every infinite-distance limit in moduli space is either an emergent string limit or a decompactification limit and suggests that each of the generator towers should correspond to either a tower of string oscillator modes or a Kaluza-Klein tower in some duality frame. In the case at hand, the winding modes are T-dual to the Kaluza-Klein modes.
\item When the bound $\lm \geq 1/\sqrt{d-2}$ is saturated in a given infinite-distance limit, it is saturated by both a tower of string oscillator modes as well as one or more towers associated with the generators of the convex hull. If, as suggested by the Emergent String Conjecture, we identify each of these generators with a Kaluza-Klein tower in some duality frame, then any infinite-distance limit either has $\lstring = \lKK = 1/\sqrt{d-2}$ or else $\sqrt{(d-1)/(d-2)} \geq \lKK >  1/\sqrt{d-2} > \lstring$. In the language of the Emergent String Conjecture, these may be considered emergent string limits or decompactification limits, respectively.
\end{itemize}

So far, these three observations may seem unique to the special case at hand. However, the Emergent String Conjecture suggests that they are rather generic in supersymmetric compactifications: any infinite-distance limit in moduli space is either an emergent string limit or a decompactification limit, and correspondingly it should feature either a tower of string oscillator modes with $\lstring = 1/\sqrt{d-2}$ or else a Kaluza-Klein tower with $\lKK > 1/\sqrt{d-2}$.

Examples in string theory and supergravity confirm these expectations \cite{Etheredge:2022opl}. Gauge couplings in supergravity in $5 \leq d \leq 9$ dimensions scale as
\begin{equation}
g_1 \sim  \exp \left( - \frac{ \kappa_d}{\sqrt{d-1}} \hat\phi -   \frac{\kappa_d }{\sqrt{(d-1)(d-2)}} \hat{\rho}  \right)  \,,~~~~g_2 \sim   \exp \left(-  \kappa_d   \sqrt{\frac{ {d-1} }{{d-2} } }\hat{\rho}  \right) 
\end{equation}
The tower Weak Gravity Conjecture then implies the existence of towers of particles with masses scaling as $m_1 \lesssim g_1$, $m_2 \lesssim g_2$. Comparison with \eqref{nKK} and \eqref{Mstringscaling} shows that these towers have precisely the scaling behavior expected of string oscillator modes and Kaluza-Klein towers, respectively.

Many string theory examples in four dimensions, as well as M-theory compactified on a torus to $d \geq 4$ dimensions, illustrate this same scaling behavior \cite{Etheredge:2022opl, More} (see also \cite{Cicoli:2017axo, Grimm:2018ohb, Corvilain:2018lgw, Heidenreich:2018kpg, Erkinger:2019umg, Joshi:2019nzi, EnriquezRojo:2020hzi}). The only exceptions come from string theory in $d=10$ dimensions, where there are no light KK modes in weakly coupled limits of string theory. In such limits, however, string oscillator modes still saturate the bound $\lstring  = 1/\sqrt{d-2}$. Meanwhile, strongly coupled limits of string theories in ten dimensions represent either weakly coupled limits of S-dual strings, whose oscillator modes satisfy $\lstring = 1/\sqrt{8}$, or else decompactification limits to 11-dimensional M-theory, whose KK modes have $\lKK = \sqrt{9/8}$.

We may thus summarize these findings as 
\begin{align}
 \sqrt{\frac{d-1}{d-2} } \geq \lm = \lKK > \frac{1}{\sqrt{d-2}} > \lstring ~~~~~\text{OR} ~~~~~
\lm = \lstring = \frac{1}{\sqrt{d-2}}  \,.
\label{summary}
\end{align}
Here, $\lKK$ and $\lstring$ should be understood respectively as the exponential decay rates of the lightest Kaluza-Klein tower and the lightest tower of string oscillator modes in the infinite-distance limit in question.

\subsubsection{Non-supersymmetric compactifications}\label{nonsusy}

Our discussion in \S\ref{toy} focused on supersymmetric compactifications, which preserve a moduli space of exactly massless scalar fields. In the remainder of this paper, however, we will we be interested primarily in scalar fields with potentials. It is natural to ask, therefore, which parts of our discussion above might carry over to scalar fields with potentials in more general, possibly non-supersymmetric string compactifications.

It is reasonable to suspect that the sharpened Distance Conjecture will apply in these cases as well. Kaluza-Klein modes, string oscillator modes, and wrapped branes are present in general compactifications of string theory, and their scaling behavior with the dilaton $\phi$ and radion $\rho$ (as in e.g.  \eqref{nKK}, \eqref{Mstringscaling}) does not rely on masslessness of these scalar fields.

\begin{figure}
\begin{center}
\center
\includegraphics[width=60mm]{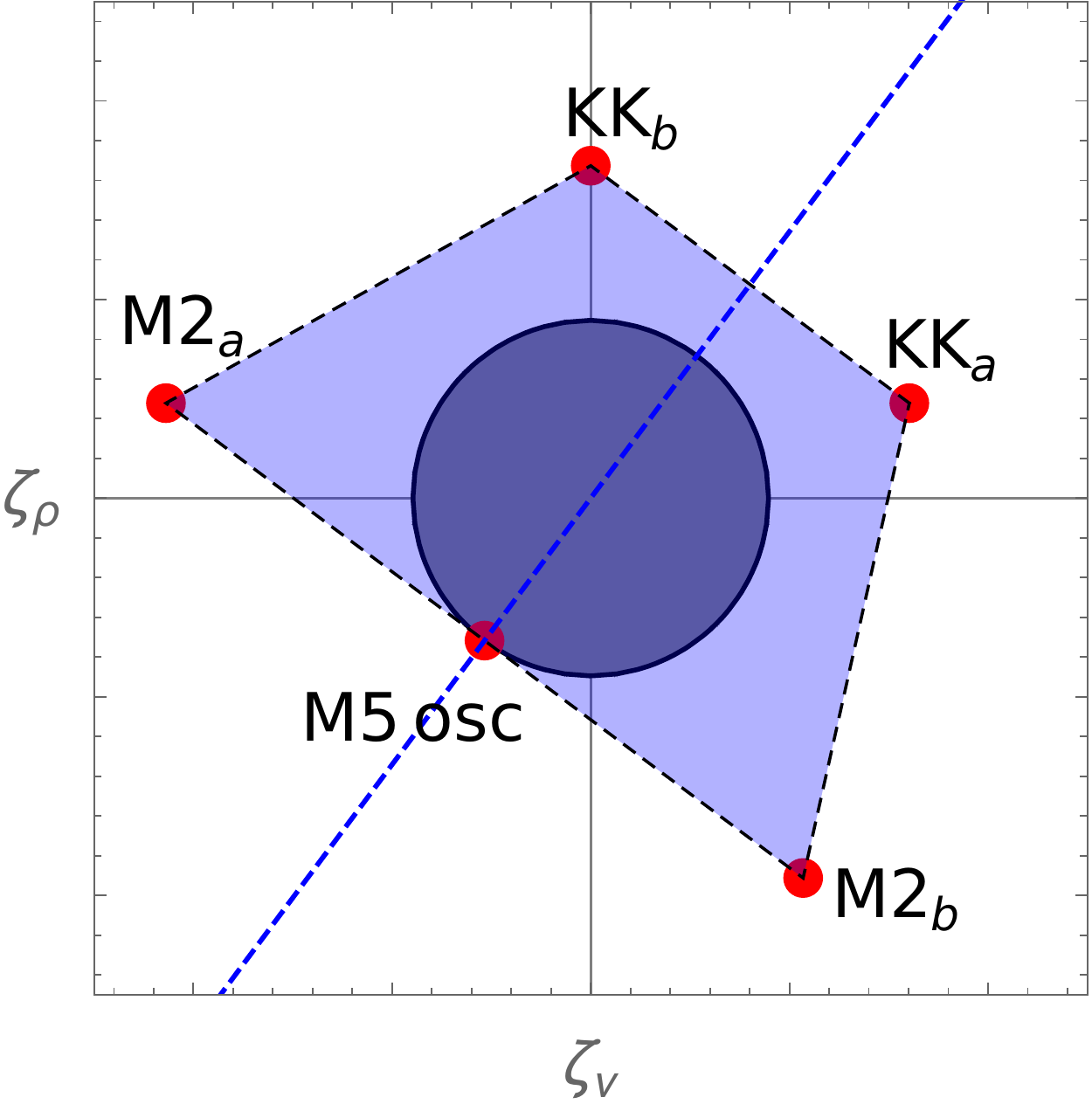}
\caption{The convex hull condition for M-theory on $\cM_a^{(2)} \times \cM_b^{(2)}$.}
\label{m2m2}
\end{center}
\end{figure}

As an example, consider M-theory compactified on a direct product of 2-manifolds, $\cM_a^{(2)} \times \cM_b^{(2)}$. Denoting the respective radion fields for the 2-manifolds by $\nu$ and $\rho$, the convex hull in $(\zeta_\nu, \zeta_\rho)$ space is generated by Kaluza-Klein modes for $\cM_a^{(2)}$, Kaluza-Klein modes for $\cM_b^{(2)}$, M2-branes wrapping $\cM_a^{(2)}$, and M2-branes wrapping $\cM_b^{(2)}$, as shown in Figure \ref{m2m2}. There are also oscillator modes associated with a string which comes from wrapping the M5-brane over $\cM_a^{(2)} \times \cM_b^{(2)}$. We see that the convex hull condition is satisfied for the two radions despite the absence of supersymmetry, and the bound $\lm \geq 1/\sqrt{d-2}$ is saturated only in the direction of field space associated with the tower of string oscillator modes.

The convex hull for M-theory on $\cM_a^{(2)} \times \cM_b^{(2)}$ shown in Figure \ref{m2m2} is precisely the same as the convex hull for M-theory compactified on an elliptically fibered K3 manifold, where $\nu$ controls the size of the base and $\rho$ controls the size of the $T^2$ fiber. This illustrates the general principle mentioned above: since the presence and scaling properties of the convex hull generators do not depend on supersymmetry, the sharpened Distance Conjecture in supersymmetric compactifications often implies the conjecture in non-supersymmetric ones.

The four generators of the convex hull in Figure \ref{m2m2} all have the length necessary to correspond to Kaluza-Klein modes in some duality frame. The points marked KK$_a$ and KK$_b$ correspond, of course, to Kaluza-Klein modes for 2-manifolds, with $|\vec \zeta| = \sqrt{7/10}$. The points marked M2$_a$ and M2$_b$ have length $|\vec \zeta| = \sqrt{6/5}$, so they have the length needed to represent Kaluza-Klein modes for decompactification limits from $d=7$ to $D=8$ dimensions.

M-theory compactified on an elliptically fibered Calabi-Yau is dual to heterotic string theory on $T^3$, and under this duality M2-brane winding modes are mapped to Kaluza-Klein modes of $S^1 \subset T^3$. Thus, in the supersymmetric case, the Emergent String Conjecture is satisfied, and every generator of the convex hull in Figure \ref{m2m2} represents Kaluza-Klein modes in some duality frame.

In the non-supersymmetric case, however, it is not clear that such a web of dualities exists. M-theory on $\cM_a^{(2)} \times \cM_b^{(2)}$ will still have the necessary M2-brane winding modes to satisfy the convex hull condition, but it is not clear that these modes correspond to Kaluza-Klein modes in any dual frame.

However, a number of works \cite{Marchesano:2019ifh, Baume:2019sry, Klaewer:2020lfg} have found evidence that in theories where the necessary dualities do not exist, the limits which naively appear to violate the Emergent String Conjecture are obstructed by quantum corrections to the scalar field space metric, which leave them at finite distance. Such obstructions may occur in massless moduli spaces (such as $\mathcal{N}=2$ supergravity in four dimensions \cite{Marchesano:2019ifh, Baume:2019sry}) or for scalar fields with potentials (as in $\mathcal{N}=1$ supergravity theories in four dimensions \cite{Klaewer:2020lfg}). If the Emergent String Conjecture indeed applies for scalar fields with potentials, then the simple Kaluza-Klein reduction analysis illustrated in Figure \ref{KKCHC} strongly suggests that the sharpened Distance Conjecture will hold as well.

We conclude that both the sharpened Distance Conjecture and the Emergent String Conjecture are likely to hold even for scalar fields with potentials.
In this context, the generators of the convex hull may or may not represent Kaluza-Klein modes in some duality frame, but the Emergent String Conjecture implies that if there is no tower of Kaluza-Klein modes or string oscillator modes with $\lambda \geq 1/\sqrt{d-2}$ in some classically infinite-distance limit, then this limit must be obstructed by quantum corrections.

For more on the sharpened Distance Conjecture in non-supersymmetric settings, see \cite{More, Basile:2022zee}.

\subsection{A Sharpened Asymptotic de Sitter Conjecture}\label{dSC}

In this subsection, we review the dimensional reduction analysis of \cite{Rudelius:2021oaz}.
We begin in $D$ dimensions with an action of the form
\begin{align}
S =  \int \rmd^D x \sqrt{-g}  \left[ \frac{1}{2 \kappa_D^2} {\cal R}_D -  \frac{1}{2} ( \nabla \hat \phi)^2 - V_D(\hat \phi) \right] \,.
 \label{eq:action}
\end{align}
We further assume that the potential $V_D(\phi)$ takes the form
\begin{equation}
V_D(\hat \phi) = V_{D,0} \exp(-  c_D \kappa_D \hat \phi  ) \,,
\end{equation}
which is typical of scalar field potentials in asymptotic limits of scalar field space in string theory \cite{Dine:1985he, Obied:2018sgi}.

Reducing to $d$ dimensions with an ansatz of the form \eqref{ansatz}, we find a potential of the form
\begin{align}
V_d &= V_D \vol_n \exp \left(-2 \kappa_d \sqrt{\frac{n}{(n+d-2)(d-2)}} \hat \rho \right)  \nonumber \\
&\sim  \exp \left(- c_D \kappa_d \hat \phi  -  2 \kappa_d \sqrt{\frac{n}{(n+d-2)(d-2)}} \hat \rho \right)  \,
\end{align}
where $\vol_n$ is the volume of the $n$-dimensional compactification space, and $\hat \rho$ is the canonically normalized radion.

Defining a new scalar field $\phi' = c_D \hat \phi  + 2 \sqrt{\frac{n}{(n+d-2)(d-2)}} \hat \rho$, this potential can be written as
\begin{align}
V_d \sim \exp \left( - c_d \kappa_d \hat \phi' \right)\,,
\label{Vred}
\end{align}
where $c_d^2 = c_D^2 + \frac{4 n}{(n+d-2)(d-2)}$. Similar to the previous subsection, this equation is solved by
\begin{equation}
c_D = \frac{2}{\sqrt{D-2}} \,,~~~~   c_d = \frac{2}{\sqrt{d-2}} \  \,.  
\end{equation}
Thus, as noted in \cite{Rudelius:2021oaz}, $c_d = 1/\sqrt{d-2}$ is singled out as a special value for the constant $c_d$, in that it is exactly preserved under dimensional reduction. This led the author to conjecture 
\begin{equation}
\frac{|\nabla V| }{ V } \geq  c_d^{\textrm{min}} \equiv \frac{2}{\sqrt{d-2}} 
\label{potbound}
\end{equation}
 in asymptotic limits of scalar field space in \cite{Rudelius:2021oaz}, building on previous work in \cite{Obied:2018sgi, Hebecker:2018vxz, Garg:2018reu, Ooguri:2018wrx, Bedroya:2019snp}.

This value of $c_d$ is also distinguished cosmologically. As shown in Appendix \ref{App}, a cosmology dominated by a scalar field rolling in a potential of the form $V_d(\hat \phi) \sim \exp( -c_d \hat \phi )$ will produce Hubble parameter $H \equiv \dot a /a$ given by
\begin{equation}
H \sim \sqrt{\dot{ \hat \phi}^2 + 2V}  \sim  \exp( - \lambda_H   \hat \phi ) \,,~~~~\lambda_H = \min \left( \frac{c_d}{2}, \sqrt{\frac{d-1}{d-2}} \right)\,.
\end{equation}
At late times, the value of $\lambda_H$ dictates the equation-of-state parameter $w$ as
\begin{equation}
w =  \min \left( -1 +  \frac{ d-2}{2(d-1)} c_d^2, 1 \right)\,.
\end{equation}
Setting $c_d = 2/\sqrt{d-2}$ thus yields
\begin{equation}
w = - \frac{d-3}{d-1} \,,
\end{equation}
which is precisely the condition for $\ddot a(t) = 0$. For $\lambda_H > 1/\sqrt{d-2}$, the universe will experience decelerated expansion $\ddot a < 0$, whereas for $\lambda_H < 1/\sqrt{d-2}$ it will undergo accelerated expansion. Imposing the bound $c_d \geq 2/\sqrt{d-2}$ is therefore equivalent to forbidding accelerated expansion at late times in asymptotic limits of scalar field space, and it is also equivalent to the strong energy condition. Several lines of evidence in favor of this bound were discussed in \cite{Rudelius:2021azq}, and we will see further evidence for it in the remainder of this paper.

\subsection{Hubble vs. Kaluza-Klein}\label{HKK}

So far, we have presented evidence in favor of the bounds $\lH \geq 1/\sqrt{d-2}$ and $\lm \geq 1/\sqrt{d-2}$ in simple Kaluza-Klein reductions. However, it remains to determine a relationship between $\lH$ and $\lm$. In this subsection, we will consider several simple examples of Kaluza-Klein reductions, and we will show that these examples obey the bound
$\lH \geq \lKK = \lm$.

\subsubsection{A dimensionally reduced potential}

Our first argument in favor of this bound again comes from Kaluza-Klein reduction of a theory with a scalar field potential $V_D$. Consider an action of the form \eqref{eq:action}, and suppose additionally that the limit $\hat \phi \rightarrow \infty$ features a tower of light particles whose masses scale as
\begin{equation}
m^{(D)} \sim \exp( - \lambda_D \hat \phi  ) \,.
\end{equation}
Suppose further that the bounds \eqref{towerbound} and \eqref{potbound} are each saturated, so $\lambda_D = c_D/2 = 1/\sqrt{D-2}$. From our discussion in \S\ref{SDC}, it is reasonable to further assume that the tower of light particles then corresponds to a tower of string oscillator modes, so $\lstring^{(D)} = 1/\sqrt{D-2}$.

After Kaluza-Klein reduction on an $n$-manifold with the ansatz \eqref{ansatz}, the potential $V_D$ descends to a potential of the form in \eqref{Vred}, with $c_d = 2/\sqrt{d-2}$. By \eqref{Mstringscaling}, the tower of string oscillator modes descends to a tower of string oscillator modes in $d=D-n$ dimensions with  
\begin{equation}
\Mstring^{(d)} \sim \exp \left( - \frac{1}{\sqrt{d-2}} \hat \phi'  \right)\,,
\label{Mshere}
\end{equation}
where crucially, the canonically normalized field $\hat \phi'$ is the same as the one in \eqref{Vred}, due to our choice $c_D =2 \lambda_D = 2/\sqrt{D-2}$.
In addition, by \eqref{nKK}, there is tower of Kaluza-Klein modes with masses which scale as
\begin{equation}
\mKK^{(d)} \sim \exp\left( - \frac{1}{\sqrt{d-2}} \hat \phi'  - \frac{1}{\sqrt{n}} \hat \rho' \right)\,,
\label{mKKhere}
\end{equation}
where $\hat \rho'$ is another canonically normalized scalar field.

At late times, the scalar field will roll along a gradient descent path $\hat \phi' \rightarrow \infty$, with $\hat \rho'$ fixed. By \eqref{Vred}, \eqref{Mshere}, and \eqref{mKKhere}, we will thus have
\begin{equation}
H \sim \sqrt{V_d} \sim \Mstring^{(d)} \sim \mKK^{(d)} \sim \exp\left( - \frac{1}{\sqrt{d-2}} \hat \phi' \right)\,,~~~\hat \phi' \rightarrow \infty \, .
\end{equation}
or
\begin{equation}
\lH = \lstring = \lKK  = \lm = \frac{1}{\sqrt{d-2}}.
\end{equation}
Thus, we see that our proposed bound $\lH \geq \lm$ is satisfied, and indeed saturated. From the analysis of the previous subsection, we already knew that the assumption $c_D = 2/\sqrt{D-2}$ will produce a potential in $d$ dimensions with $\lambda_H = 1/\sqrt{d-2}$, and by \eqref{summary} we further expect $\lm \geq 1/\sqrt{d-2}$. The only way to satisfy $\lH \geq \lm$, therefore, is if $\lm = 1/\sqrt{d-2}$, which is precisely what we have found.

Graphically, this is possible because gradient descent pushes the scalar field along the $\hat \phi'$ direction in field space, which is the direction in scalar field space corresponding to the red dot labeled $\vec{\zeta}_{\textrm{part}}$ in Figure \ref{KKCHC}. This is the unique point along the convex hull with $\lm = 1/\sqrt{d-2}$.

\subsubsection{A potential from internal curvature}\label{internal}

Even if the $D$-dimensional theory has no scalar field potential ($V_D = 0$), a potential can be generated after compactification to $d=D-n$ dimensions if the compactification manifold has a nonzero scalar curvature $\mathcal{R}_n$. The potential asymptotically takes the form
\begin{equation}
V_d \sim -  \mathcal{R}_n \exp\left(  - 2 \sqrt{\frac{n+d-2}{n(d-2)}}  \hat \rho \right)\,.
\label{internalV}
\end{equation}
For $\mathcal{R}_n < 0$, this yields $\lambda_H = \sqrt{\frac{n+d-2}{n(d-2)}} $, and it means that at late times, the scalar field will roll along the gradient descent path $\hat \rho \rightarrow \infty$, with $\hat \phi$ fixed. Comparing with Figure \ref{KKCHC}, we see that gradient descent pushes the scalar field along the $\hat \rho$ direction in field space, which is the direction in scalar field space corresponding to the red dot labeled $\vec{\zeta}_{\textrm{KK}}$.

Along this direction in field space, a tower of Kaluza-Klein modes will also become light. From \eqref{nKK}, we see that in the $\hat \phi \rightarrow \infty$ limit, these behave as
\begin{equation}
m_{\textrm{KK}}^{(d)} \sim \exp \left(-  \kappa_d   \sqrt{\frac{ {n+d-2} }{{n(d-2)} } }\hat{\rho}  \right) \,.
\end{equation}
Thus, we have
\begin{equation}
\lKK = \sqrt{\frac{n+d-2}{n(d-2)}} \,,
\end{equation}
which yields $\sqrt{(d-1)/(d-2)} \geq \lH = \lKK = \lm \geq 1/\sqrt{d-2}$, as expected. 

It is worth noting that the compactification manifold breaks supersymmetry for $\mathcal{R}_n < 0 $. As discussed in the \S\ref{nonsusy}, the evidence for the Emergent String Conjecture is not as firm in non-supersymmetric settings as it is for supersymmetric theories, so it is not clear that every infinite-distance limit is either a decompactification limit or an emergent string limit. However, for the purposes of late-time cosmology, the relevant infinite-distance limit in scalar field space is the gradient flow direction of the potential $\hat \rho \rightarrow \infty$, and we see here that this limit is indeed a decompactification limit.

\subsubsection{Dimension reduction of a string-frame cosmological constant}\label{ssec:SF}

Consider a string-frame action in $D$ dimensions of the form
\begin{equation}
S = \frac{1}{2}  \int d^Dx \sqrt{-g} \left[ e^{-2 \phi} \left( \mathcal{R}_D - \frac{1}{2} H_3 \wedge \star H_3 + 4 \partial_\mu \partial^\mu \phi \right) - 2 \Lambda_D \right] \,,
\end{equation}
where $\Lambda_D >0$ is a cosmological constant and $H_3 = dB_2$. After converting to Einstein frame, this yields a potential which decays exponentially in the weak coupling limit $\hat \phi \rightarrow \infty$ as
\begin{equation}
V(\phi) \sim \Lambda_D \exp\left( -\frac{D}{\sqrt{D-2}}  \hat \phi \right)\,.
\end{equation}
This limit also features a tower of light oscillator modes for the string charged under $B_2$, whose masses scale as
\begin{equation}
\Mstring \sim \exp\left( -\frac{1}{\sqrt{D-2}} \hat \phi \right)\,.
\end{equation}
By the analysis in Appendix \ref{App}, $\lH =  \sqrt{(D-1)/(D-2)} $ since $c_D > 2  \sqrt{(D-1)/(D-2)} $. Thus, we have $\lH =  \sqrt{(D-1)/(D-2)} > \lm = \lstring = 1/\sqrt{D-2}$ in $D$ dimensions.

After reduction to $d=D-1$ dimensions, the potential picks up an additional dependence on the radion $\hat \rho$, so at the classical level,
\begin{equation}
V_{\text{cl}}(\hat \phi, \hat \rho) \sim \exp\left( -\frac{d+1}{\sqrt{d-1}}  \hat \phi -  \frac{2}{\sqrt{(d-1)(d-2)}}  \hat \rho  \right) \,.
\end{equation}
In addition, there is a one-loop Casimir energy contribution to the potential from light particles of the form \cite{ArkaniHamed:2007gg}:\footnote{We are thankful to Ben Heidenreich for pointing out the relevance of the Casimir energy in this example.}
\begin{equation}
V_{\text{csmr}}(\lambda) = \mp \frac{2}{(2 \pi R)^{d} \Omega_{d}} \zeta(d+1) \exp\left( - d \sqrt{\frac{d-1}{d-2}} \hat \rho \right)  \,,~~~\Omega_d = \frac{2 \pi^{(d+1)/2}}{\Gamma(\frac{d+1}{2})}\,.
\label{VC}
\end{equation}
Here, $R$ is the radius of the circle, $\zeta(x)$ is the Riemann zeta function, $\Omega_{d}$ is the volume of the unit $d$-sphere, and the $-$ sign is for bosons or fermions with antiperiodic boundary conditions, while the $+$ sign is for fermions with periodic boundary conditions. The 1-loop Casimir energy for a particle heavier than $1/R$ is exponentially suppressed as $\e^{-2 \pi m R}$, so at low energies and for $R$ large, the relevant contributions come only from massless particles, and these will be positive if periodic fermions outnumber bosons and antiperiodic fermions.

If massless periodic fermions indeed outnumber bosons and antiperiodic fermions, the Casimir energy will then contribute to the potential with a positive sign as
\begin{equation}
V_{\text{csmr}}(\hat \phi, \hat \rho) \sim  \exp\left( - d \sqrt{\frac{d-1}{d-2}} \hat \rho \right)  \,.
\label{Casimir}
\end{equation}
The combined potential then takes the form
\begin{equation}
V(\hat \phi, \hat \rho) =  V_{\text{cl},0}  \exp\left( -\frac{d+1}{\sqrt{d-1}}  \hat \phi -  \frac{2}{\sqrt{(d-1)(d-2)}}  \hat \rho  \right) +  V_{\text{csmr},0}  \exp\left( - d \sqrt{\frac{d-1}{d-2}} \hat \rho \right)  \,.
\label{Vss}
\end{equation}
The gradient of this potential satisfies the bound
\begin{equation}
\frac{|\nabla V|}{V} \geq \frac{d}{\sqrt{d-2}}\,,
\end{equation}
saturating the bound in the limit $\hat \rho, \hat \phi \rightarrow \infty$ with $ \hat \rho / \hat \phi = 1/ \sqrt{d-2} $.\footnote{It is interesting to note that the potential in $D$ dimensions satisfies $|\nabla V| /V = D/\sqrt{D-2}$, so in this sense the coefficient $c_d = d/\sqrt{d-2}$ is exactly preserved under dimensional reduction in the presence of a positive Casimir energy.} Since $|\nabla V|/V > 2\sqrt{(d-1)/(d-2)}$ everywhere in scalar field space, the kinetic energy of the rolling field necessarily dominates over the potential energy at late times, so $\lH = \sqrt{(d-1)/(d-2)}$ by the analysis of Appendix \ref{App}.

The fact that the kinetic energy of the field dominates over its potential energy further implies that Hubble friction will not be strong enough to slow the rolling field and force it along the path of gradient descent.
In this case, the motion of the field at late times may be determined by either one of these exponential terms or by a combination thereof, depending on initial conditions. The field may roll off to infinity in a range of directions in the $\hat\phi$-$\hat\rho$ plane, provided the constraints
\be
\frac{|V'|}{V} \geq 2 \sqrt{\frac{d-1}{d-2}} \,,~~~~V' < 0
\label{dirconst}
\ee
 are satisfied, where $V'$ denotes the directional derivative of $V$ in the direction of interest. For the potential in \eqref{Vss}, these constraints are satisfied in the infinite-distance limits with $\hat\phi, \hat \rho \rightarrow \infty$ provided
\begin{equation}
\frac{2}{\sqrt{(d-2)(d+2)}} \hat \phi \leq \hat \rho \leq \frac{d(d+1) +2}{2d \sqrt{d-2}}  \hat \phi \,.
\label{const1}
\end{equation}
There is also a constraint
\begin{equation}
\frac{1}{\sqrt{d-2}} \hat \phi \leq \hat \rho\,,
\label{const2}
\end{equation}
which ensures that the Kaluza-Klein scale remains below the string scale at late times. If this is violated, the effective field theory used to compute the potential in \eqref{Vss} breaks down, and further analysis is needed.

Meanwhile, there are towers of Kaluza-Klein modes, winding modes, and string oscillator modes whose masses scale as 
\begin{align}
\mKK \sim  \exp\left( -\sqrt{ \frac{ d-1}{ d-2 } } \hat \rho \right) \,, & ~~~ m_{\textrm{wind}} \sim\exp\left( - \frac{2}{\sqrt{d-1}}  \hat \phi  + \frac{d-3}{\sqrt{(d-1)(d-2)}} \hat \rho \right)\,, \nonumber \\
 \Mstring & \sim  \exp\left(  -\frac{1}{\sqrt{d-1}} \hat \phi - \frac{1}{\sqrt{(d-1)(d-2)}} \hat \rho \right)\,.
\end{align}
Taking $\hat \phi, \hat \rho \rightarrow \infty$ subject to the constraints in \eqref{const1} and \eqref{const2}, the scaling coefficients satisfy
\begin{equation}
\sqrt{\frac{d-1}{d-2}} = \lH \geq \lm =  \lKK  \geq \frac{1}{\sqrt{d-2}} \geq \lstring \geq \lambda_{\text{wind}} \,,
\end{equation}
as expected.

\subsubsection{Supersymmetry and a vanishing potential}

A supersymmetric compactification which preserves at least eight supercharges will leave a moduli space parametrized by vevs of massless scalar fields. As discussed in \S\ref{SDC} above, the combination of the sharpened Distance Conjecture and the Emergent String Conjecture suggest that the lightest tower in any direction in moduli space should be either a tower of Kaluza-Klein modes with $\sqrt{(d-1)/(d-2)} \geq \lKK \geq 1/\sqrt{d-2}$ or a tower of string oscillator modes with $ \lstring = 1/\sqrt{d-2}$. 

To study the behavior of the Hubble scale in this scenario, let us suppose that a canonically normalized massless modulus $\phi$ in $d$ spacetime dimensions is given some initial kick of kinetic energy, $\dot \phi (t=0) = v$, setting $v>0$ without loss of generality. (We omit the $\hat \cdot$ for ease of notation.) The evolution of $\phi$ with time is then given by solving the equation of motion (see e.g. \cite{Hellerman:2006nx})
\begin{equation}
\ddot \phi + (d-1) H \dot \phi = 0\,,~~~~H^2 = \frac{ \dot \phi^2}{(d-1)(d-2)}\,,
\end{equation}
where we have set $V(\phi) = 0$ and supposed that there are no additional contributions to the energy density from radiation or matter.\footnote{As noted in \cite{Ooguri:2006in}, if there are contributions to the energy density aside from the scalar field, then Hubble friction will prevent the massless scalar field from accessing the asymptotic regimes of moduli space, as the field will traverse only a finite distance in the course of its evolution.} Setting $\dot \phi (t=0) = v$, this pair of equations is solved by
\begin{equation}
\phi(t) = \phi_0 + \sqrt{\frac{d-2}{d-1} } \log \left( 1 + \sqrt{\frac{d-1}{d-2}} v t \right)\,.
\end{equation}
In addition, the Hubble parameter $H$ scales with time as $H \sim 1/t$, which by the equation above means that $H$ scales with $\phi$ as
\begin{equation}
H \sim \exp \left( - \sqrt{\frac{d-1}{d-2}}  \phi \right)
\end{equation}
in the limit $\phi \rightarrow \infty$. Thus we have $\lH = \sqrt{(d-1)/(d-2)}$ in the case of a vanishing potential, $V(\phi) = 0$, in agreement with the analysis of Appendix \ref{App} for $c = \infty$.

Remarkably, this is precisely the condition needed to ensure $\lH \geq \lm$, since by \eqref{summary} we have $\lm  \leq  \sqrt{(d-1)/(d-2)}$ in all asymptotic limits of moduli space, with saturation $ \lm = \lKK =  \sqrt{(d-1)/(d-2)}$ occurring only if $\phi$ is the radion for a decompactification from $d$ dimensions to $D=d+1$ dimensions. In other directions in moduli space, the inequality will be strict, $\lH > \lm$.

Note that in this special case of an exact moduli space with a vanishing potential, the bound $\lH \geq \lm $ applies to \emph{any} infinite-distance limit. This is due to the fact that, in the case of a vanishing potential, a field given an initial kick in some direction of moduli space will continue in this direction (i.e., along a geodesic) for all time. In contrast, in the presence of a potential, only certain infinite-distance limits in moduli space will be dynamically accessible to the scalar field at late times, as the field cannot roll uphill indefinitely.


\section{String Theory}\label{sec:ST}

In this section, we consider a number of examples in string theory, all of which satisfy our conjectured bounds $ \lH \geq \lm  \geq 1/\sqrt{d-2}$ at late times in asymptotic limits of scalar field space.

\subsection{Supercritical String Theory}

Our first example is supercritical string theory in $d > 10$ spacetime dimensions. This theory has a potential of the form \cite{Hellerman:2006nx}
\begin{equation}
V \sim \exp \left( - \frac{2}{\sqrt{d-2}} \hat \phi  \right) \,,
\end{equation}
where $\hat \phi$ is the canonically normalized dilaton. The limit $\hat \phi \rightarrow \infty$ is a weak coupling limit, which introduces a tower of string oscillator modes with 
\begin{equation}
\Mstring \sim \exp \left( - \frac{1}{\sqrt{d-2}} \hat \phi  \right) \,.
\end{equation}
Thus we have $\lH = \lstring = \lm = 1/\sqrt{d-2}$, saturating our proposed bounds.\footnote{Reference \cite{Dodelson:2013iba} constructed solutions of supercritical string theories with a finite period of accelerated cosmological expansion inside a finite region in scalar field space. This does not contradict the claims of this paper, however, since we are concerned solely with late-time cosmology in asymptotic limits of scalar field space.}

\subsection{$O(16) \times O(16)$ Heterotic String Theory}

Our next example is $O(16) \times O(16)$ string theory: the unique tachyon-free non-supersymmetric heterotic string theory \cite{Alvarez-Gaume:1986ghj}. This theory has a string frame cosmological constant, so it falls under the general analysis of \S\ref{ssec:SF}. In particular,
after converting to Einstein frame, we find a potential which decays exponentially in the weak coupling limit $\hat \phi \rightarrow \infty$ as \cite{Obied:2018sgi}
\begin{equation}
V(\phi) \sim \exp\left( -\frac{5}{\sqrt{2}}  \hat \phi \right)\,,
\end{equation}
which implies that the kinetic energy dominates the potential energy of the rolling field at late times, and $\lH = \sqrt{9/8}$. This limit also features a tower of light string oscillator modes whose masses scale as
\begin{equation}
\Mstring \sim \exp\left( -\frac{1}{\sqrt{8}} \hat \phi \right)\,,
\end{equation}
so $ \lstring  = 1/\sqrt{8}$. One may view string states charged under $O(16) \times O(16)$ as Kaluza-Klein modes associated with the extra 16 dimensions of the left-moving bosonic string, and these scale with $\phi$ in the same fashion, so $\lKK = 1/\sqrt{8}$. As expected, $\lH > \lKK = \lstring = \lm = 1/\sqrt{d-2}$.

Upon compactifying this theory on a circle to $d=9$ dimensions, the potential takes the form
\begin{equation}
V(\hat \phi, \hat \rho) = V_{\text{cl}} + V_{\text{csmr}}  = V_{\text{cl},0}  \exp\left( -\frac{5}{\sqrt{2}}  \hat \phi -  \frac{1}{\sqrt{14}}  \hat \rho  \right) +  V_{\text{csmr},0}  \exp\left( - 9 \sqrt{\frac{8}{7}} \hat \rho \right)  \,.
\end{equation}
Here, the sign of the Casimir energy is positive provided we choose periodic boundary conditions for the fermions, as fermions outnumber bosons at the massless level of $O(16) \times O(16)$ heterotic string theory (see e.g. \S11.3 of \cite{Polchinski:1998rr}).

Following the general analysis of \S\ref{ssec:SF} above, we have
\begin{equation} 
\sqrt{\frac{d-1}{d-2}} = \lH \geq \lm =  \lKK  \geq \frac{1}{\sqrt{d-2}} \geq \lstring \geq \l_{\text{wind}} \,. 
\end{equation}
with $d=9$.

\subsection{Type II Compactifications}

In this subsection, we consider compactifications of 10d Type IIA/IIB supergravity with D$p$-brane and O$p$-plane sources to four dimensions, following \cite{Andriot:2020lea}. The 10d metric takes the form
\begin{equation}
d s_{10}^2 = d s_4^2 + ds_6^2 \,,~~~~ds_4^2 = g_{\mu\nu}(x) dx^\mu dx^\nu\,,~~~~ds_6^2 = h_{mn}(y) dy^m dy^n\,.
\end{equation}
The $p$-brane and $p$-plane sources fill the four-dimensional spacetime and wrap $p-3$ directions in the 6d compactification space. This compactification space can then be split into directions parallel to the sources and those perpendicular to the sources, with
\begin{equation}
ds_6^2 = \rho \left( \sigma^{p-9} ds_{\parallel}^2 + \sigma^{p-3} ds_\perp^2 \right) \,, 
\end{equation}
where $\rho$ and $\sigma$ are metric fluctuations, which correspond to scalar fields in 4d. The 4d potential also depends on a dilaton $\tau$, which is a function of the 10d dilaton $\phi$ and the radion $\rho$.
After freezing $\sigma =1$, the potential can be written in terms of the scalar fields $\rho$, $\tau$ as \cite{Hertzberg:2007wc, Andriot:2019wrs, Andriot:2020lea}
\begin{align}
V = - \frac{1}{2} \tau^{-2} \left( \rho^{-1} \mathcal{R}_6 - \frac{1}{2} \rho^{-3} |H|^2 \right) - \frac{g_s}{2} \tau^{-3} \rho^{\frac{p-6}{2}} \frac{T_{10}}{p+1} + \frac{1}{4} g_s^2 \tau^{-4} \sum_{q=0}^6 \rho^{3-q} |F_q|^2 \,.
\label{VII}
\end{align}
Here, $H$ and $F_q$ are fluxes, $g_s$ is the string coupling, $\mathcal{R}_6 $ is the scalar curvature of the compact dimensions, and $T_{10}$ is a complicated function of fluxes and scalars which can be found in \cite{Andriot:2016xvq, Andriot:2020lea}.
The fields $\tau$, $\rho$ are related to the canonically normalized 4d dilaton $\hat \tau$ and radion $\hat \rho$ by
\be
\hat \tau = \sqrt{2} \log \tau \,,~~~~\hat \rho = \sqrt{\frac 3 2} \log \rho\,.
\ee
With this, we may first recognize the term $ \rho^{-1} \mathcal{R}_6$ as the portion of the potential generated by internal curvature, as discussed previously in \S\ref{internal}. This portion of the potential may be written as
\begin{equation}
V_{\text{int}} =  - \frac{1}{2} \tau^{-2} \rho^{-1} \mathcal{R}_6 =   - \frac{1}{2}  \exp\left( - \sqrt{2} \hat \tau - \sqrt{2/3} \hat \rho \right) \mathcal{R}_6 ~~  \Rightarrow ~~ \frac{|\nabla V_{\text{int}} |}{|V_{\text{int}}| } = 2 \sqrt{2/ 3}\,.
\end{equation}
This matches with \eqref{internalV} for a compactification from $D=10$ to $d=D-n=4$ dimensions with a potential generated by internal curvature. Thus, if the potential is dominated asymptotically by the contribution from internal curvature, we will have $\lH = \sqrt{2/3} = \lKK$.

More generally, we expect that the potential will be dominated asymptotically by a single term in the potential \eqref{VII}, which scales as $V \sim \tau^{-n} \rho^{-l}$ for some coefficients $n$, $l$. The coefficient $\lH$ is then given asymptotically by 
\begin{equation}
\lH = \min \left( \frac{1}{2} \frac{|\nabla V|}{V},\sqrt{\frac{3}{2}} \right) = \min \left(  \frac{1}{2} \sqrt{ \frac{n^2}{2} +  \frac{2 l^2}{3} },\sqrt{\frac{3}{2}} \right) \,.
\label{lHII}
\end{equation}
If $|\nabla V| /V \leq 2 \sqrt{3/2}$, the direction of the field at late times is the gradient flow direction of the potential. If $|\nabla V| /V > 2 \sqrt{3/2}$, however, then the field may roll at late times in any direction in the $\hat \tau$-$\hat \rho$ plane that satisfies the constraint \eqref{dirconst}, as the potential is negligible from the perspective of the late-time dynamics by the analysis of Appendix \ref{App}.

The fact that every term in the potential scales has $n \geq 2$ immediately implies $\lH \geq 1/\sqrt{2}$, satisfying our proposed bound against accelerated expansion, $\lH \geq 1/\sqrt{d-2}$. Note that this is a stronger bound on the gradient of the potential than those discussed in e.g. \cite{Flauger:2008ad, Andriot:2016xvq, Andriot:2020lea}, simply because we are focusing our attention on asymptotic regimes of scalar field space, whereas most previous works have derived bounds that apply everywhere in field space (see however \cite{Cicoli:2021fsd}, which also found accelerated expansion to be impossible in asymptotic regimes of scalar field space).

Meanwhile, the 10d Kaluza-Klein mass scale $\mKK$ scales with the moduli as $\mKK \sim \tau^{-2} \rho^{-1}$.
This means that the coefficient $\lKK$ for the 10d Kaluza-Klein scale is bounded above as $\lKK \leq \sqrt{2/3}$ in any direction in scalar field space, and
in the gradient flow direction for the potential $V \sim \tau^{-n} \rho^{-l}$ it is given by
\begin{equation}
\lKK =  \frac{1}{ \sqrt{ \frac{n^2}{2} + \frac{2 l^2}{3}   }   } \left( \frac{n}{2} + \frac{l}{3}  \right) \,.
\label{lKKII}
\end{equation}
For $l \geq 0$, we expect $\lm = \lKK$, as the Kaluza-Klein modes remains lighter than the string modes, which in turn remain lighter than the string winding modes.

Comparing with \eqref{lHII}, we see that for $n \geq 3$, we have $\lH > \lKK $ for all real $l$.
For $n=2$, our proposed bound $\lH \geq \lKK$ requires $l \leq 0$ or $l \geq 1$. The terms in the potential \eqref{VII} with $n<3$ have $n=2$ and $l = 1$ or $l=3$, so the bound $\lH \geq \lKK$ is satisfied no matter which of the terms in the potential dominates. Note that a dominant term of the form $V \sim \tau^{-2} \rho^{-l}$ with $0 <l < 1$ would present a counterexample to our proposed bound $\lH \geq \lm$, so the absence of such a term in \eqref{VII} provides evidence in favor of our proposal.

\subsection{Heterotic Compactifications}

Heterotic compactifications are similar to Type II compactifications. At order $(\alpha')^0$, the potential is given by the NS-NS part of \eqref{VII} \cite{Andriot:2020lea}, namely
\begin{align}
V = - \frac{1}{2} \tau^{-2} \left( \rho^{-1} \mathcal{R}_6 - \frac{1}{2} \rho^{-3} |H|^2 \right)  \,,
\label{Vhet}
\end{align}
where the canonically normalized fields are again given by
\be
\hat \tau = \sqrt{2} \log \tau \,,~~~~\hat \rho = \sqrt{\frac 3 2} \log \rho\,.
\ee
From here, the analysis is identical to the Type II case above. If the first term dominates with $\mathcal{R}_6 < 0$, then at late times, the Hubble constant scales with proper field distance as $\lH  =\lm  = \lKK  = \sqrt{2/3} $. If this term is absent, and the $|H|^2$ term dominates, then by \eqref{lHII} and \eqref{lKKII} we will instead have $\lH = \sqrt{3/2}$ and $\lKK = \lm= 1/\sqrt{2}$. In either case, we indeed have $\lH \geq \lm = \lKK \geq 1/\sqrt{d-2}$.

At order $\alpha'$, the dilaton couples universally to the potential with a factor of $\tau^{-2}$ \cite{Gautason:2012tb}. Setting $n=2$ in \eqref{lHII}, this ensures $\lH \geq 1/\sqrt{2} = 1/\sqrt{d-2}$, and provided the dominant term takes the form $V \sim \tau^{-2} \rho^{-l}$ with $l  = 0$ or $l \geq 1$, \eqref{lKKII} ensures $\lH \geq \lm = \lKK$, as well.


\section{Asymptotic Cosmology}\label{sec:COSMO}

In this paper, we have argued for the constraint $\lH \geq \lm \geq 1/\sqrt{d-2}$ within asymptotic limits of scalar field space in quantum gravity. In this section, we examine the cosmological consequences of this constraint, and we argue that it is not merely a numerical accident, but rather a signpost pointing towards the possible late-time behavior of cosmological solutions in the landscape.

To begin, consider the bound $\lH \geq \lm$. This bound ensures that the Hubble scale remains parametrically at or above the mass scale of the lightest tower of particles. A number of recent works \cite{Bedroya:2019snp, Agrawal:2019dlm}, motivated by the Distance Conjecture, have entertained the possibility of a tower of particles whose masses are exponentially lighter than the Hubble scale, but our arguments in this paper suggest that this possibility does not actually occur in asymptotic limits of scalar field space in string theory.

The bound $\lH \geq \lm$ further implies that the Hubble scale $H$ remains parametrically at or below the Kaluza-Klein scale $\mKK = 1/L$ and the string scale $\Mstring$, since each of these scales feature a tower of light particles. The former condition, $H \leq \mKK$, means that the horizon size is larger than the scale of the extra dimensions $L$, which ensures that the cosmology can be viewed, at low energies, as a $d$-dimensional FRW cosmology. The latter condition, $H < \Mstring$, implies that the theory at low energies $E \sim H$ can be viewed as a low-energy effective field theory rather than a string theory with a Hagedorn density of states.

Next, consider the bound $\lH \geq 1/\sqrt{d-2}$. A scalar field rolling down the potential $V \sim \exp(- 2 \lH \phi )$ gives rise to an equation of state parameter \cite{Obied:2018sgi, Agrawal:2018own}
\begin{equation}
w \equiv p/\rho = -1 +  \frac{2 (d-2)}{d-1} \lH^2 \,.
\end{equation}
The bound $\lH \geq 1/\sqrt{d-2}$ thus implies $w \geq -(d-3)/(d-1)$, which is precisely the strong energy condition in $d$ dimensions, which is in turn the condition that forbids accelerated expansion of the universe, $\ddot a \leq 0$.

In sum, then, the bounds $\lH \geq \lm \geq 1/\sqrt{d-2}$ ensure that (a) the Hubble scale remains at or below the Kaluza-Klein scale and the string scale in the asymptotic future and (b) the universe does not undergo indefinite accelerated expansion. These conditions, in turn, have important consequences for the asymptotic structure of the universe, as we will now explain.

\subsection{Asymptotic Structure}

Consider a $D=d+n$-dimensional spacetime comprised of a $d$-dimensional flat expanding FRW universe and an $n$-dimensional compactification space with metric 
\begin{equation}
d s^2 = -dt^2 + a(t)^2 [ dr^2  + r^2 d \Omega_{d-2}^2 ] + b(t)^2 h_{lm} dy^l dy^m\,,
\end{equation}
where, $l, m = 1,...,n$. Assuming the evolution of the $d$-dimensional spacetime is that of a perfect fluid with equation of state parameter $w$, the scale factor $a$ will grow with time as
\begin{equation}
a(t) \sim t^{\frac{2}{(w+1)(d-1)}}\,.
\end{equation}
A radial null geodesic has $d t = a d r$, or $
r \sim t^{1 - \frac{2}{(w+1)(d-1)} } $.
This means that far out along the null geodesic, the length scale of the asymptotic $S^{d-2}$ grows as
\begin{equation}
L_{S^{d-2}} = a(t) r(t) \sim t \,,
\end{equation}
notably irrespective of $w$.

\begin{figure}
\begin{center}
\begin{subfigure}{0.3\textwidth}
\center
\includegraphics[width=37mm]{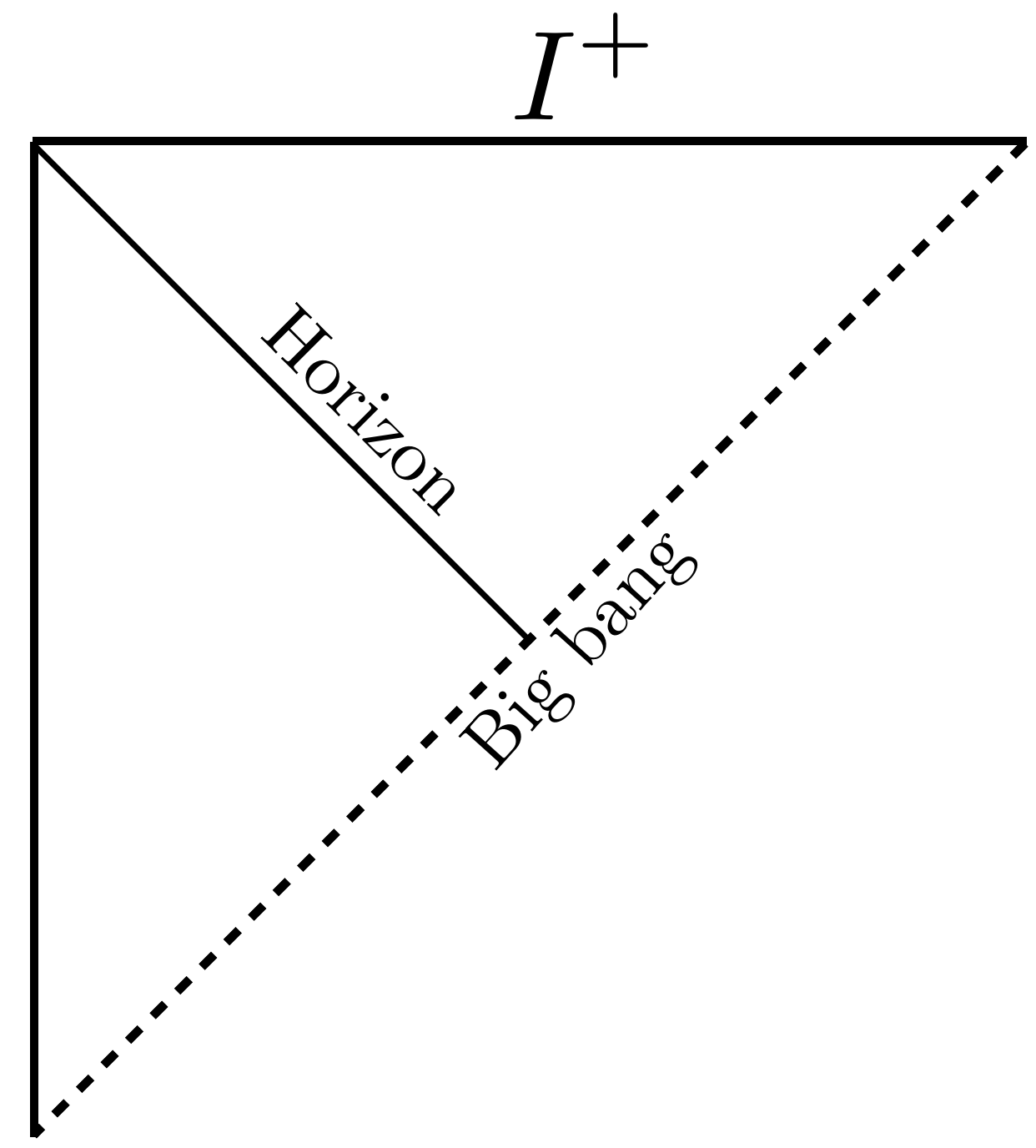}
\caption{$\ddot a > 0$}
\end{subfigure}
\begin{subfigure}{0.3\textwidth}
\center
\includegraphics[width=30mm]{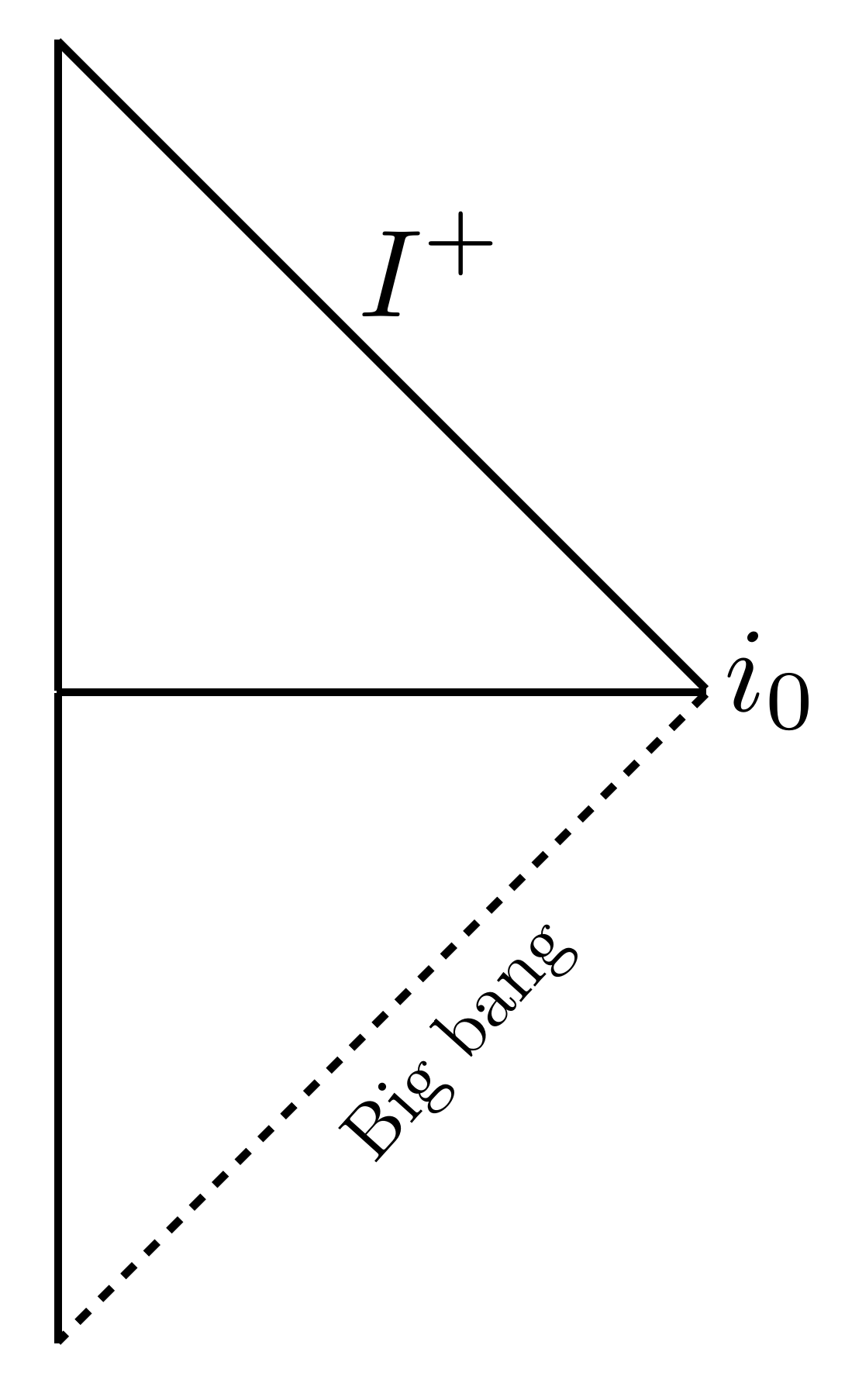}
\caption{ $\ddot a = 0$}
\end{subfigure}
\begin{subfigure}{0.3\textwidth}
\center
\includegraphics[width=40mm]{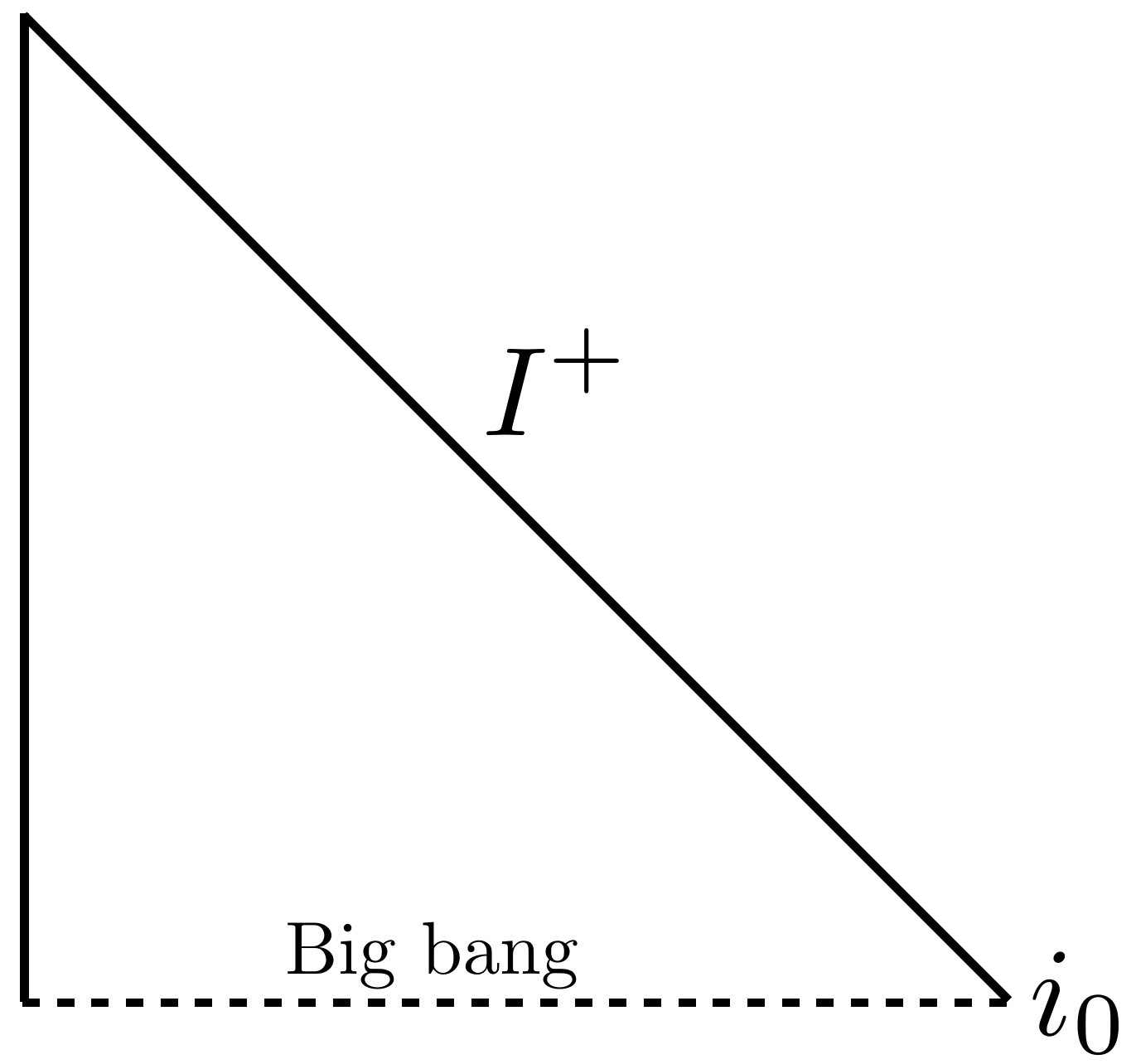}
\caption{$\ddot a < 0$} 
\end{subfigure}
\caption{Penrose diagrams for flat FRW cosmologies with $\ddot a >0$, $\ddot a = 0$, and $\ddot a <0$. Figures adapted from \cite{Bousso:2004tv, Hellerman:2006nx, Rudelius:2021azq}.}
 \label{Penrose}
\end{center}
\end{figure}

Meanwhile, the Hubble parameter is given by $H = \dot a/a = 2/[(w+1)(d-1)t]$. Our rolling solution has $H \sim \exp( - \lH \phi )$, so $
t \sim \exp( \lH \phi ) $.
Meanwhile, the characteristic length scale of the $n$-dimensional compactification space grows as
\begin{equation}
L_{\textrm{KK}} \sim 1/ \mKK \sim \exp( \lKK \phi) \sim t^{\lKK/\lH}\,.
\end{equation}
Thus, the ratio of the size of the asymptotic $S^{d-2}$ to the size of the $n$-dimensional compactification space grows as
\begin{equation}
\frac{L_{S^{d-2}} }{ L_{\textrm{KK}}} \sim t^{1 - \lKK/\lH }\,.
\end{equation}
For $\lKK < \lH$, therefore, the relative size of the compact space to the asymptotic $S^{d-2}$ tends to zero at null infinity. This is analogous to what happens at spatial infinity in the AdS/CFT correspondence for AdS$_{d+1} \times S^{n}$: the size of the compact $S^n$ remains finite as $r \rightarrow \infty$, while the size of the $d$-dimensional space diverges. The asymptotic symmetry group then acts only in AdS$_{d+1}$ as opposed to $S^n$, and the AdS boundary correlators are controlled by a CFT$_d$ as opposed to a CFT$_{d+n}$.

In the case at hand, the bound $\lH \geq 1/\sqrt{d-2}$ ensures that the $d$-dimensional FRW spacetime does, in fact, feature an asymptotic boundary at null infinity \cite{Hellerman:2001yi}, as can be seen by comparing the Penrose diagrams in Figure \ref{Penrose}. When $\lH < 1/\sqrt{d-2}$, on the other hand, there is a future horizon, and the only asymptotic boundary lies at future spacelike infinity. The bound $\lH > \lKK \geq 1/\sqrt{d-2}$ then implies that the asymptotic symmetries of the theory act only in the $d$-dimensional spacetime, as the size of the compact $n$-manifold shrinks indefinitely relative to the size of the $(d-2)$-sphere at null infinity.

 Things are a bit more complicated when $\lH = \lKK$. In this case, the relative size of the $n$-manifold is not exponentially suppressed with respect to $\hat \phi$ relative to the size of the $(d-2)$-sphere. There may still be a power-law suppression with $\hat \phi$, ensuring that ${L_{S^{d-2}} }  /  { L_{\textrm{KK}}} \rightarrow \infty$ as $t \rightarrow \infty$. Or, it may be that ${L_{S^{d-2}} }  /  { L_{\textrm{KK}}}$ approaches a finite constant at late times, in which case the boundary at null infinity is effectively $n+d-2$-dimensional. We leave a more thorough investigation of this case for future research.

\subsection{Stability of Asymptotic Quintessence}

Above, we have shown that the bound $\lH \geq 1/\sqrt{d-2}$ ensures that indefinite accelerated expansion, or quintessence, cannot occur in asymptotic regimes of scalar field space. It is tempting to turn this argument around and view the absence of indefinite accelerated expansion as an explanation for why the bound $\lH \geq 1/\sqrt{d-2}$ seems to hold in asymptotic limits of scalar field space in string theory. For this explanation to work, however, we must further argue that a violation of the bound $\lH \geq 1/\sqrt{d-2}$ would lead to stable quintessence, which persists in the asymptotic limit $t \rightarrow \infty$.

To this end, let us specialize to the case of $d=4$ dimensions and consider a scalar field rolling in a potential $V =V_0 \exp(- 2 \lH \phi)$, with $\lH < 1/\sqrt{2}$. This gives rise to a universe with quintessence, also known as ``Q-space,'' with equation of state parameter
\be
w  = -1 +  \frac{4}{3} \lH^2 < - \frac{1}{3} \,.
\ee

The thermodynamics of Q-space is similar to that of metastable de Sitter space with a time-dependent Hubble parameter \cite{Bousso:2004tv}. This comparison is most rigorous when the equation of state parameter is very close to $-1$, i.e., $0 < w +1 \ll 1$, ensuring that the Hubble parameter changes slowly relative to the Hubble size of the system, $|\dot H|/H^2 = 3(w+1)/2 \ll 1$. In Q-space, a comoving observer will see an apparent horizon of radius $R_A = 3 t(w+1)/2 $ with a Bekenstein-Hawking entropy
\begin{equation}
S_{\text{Q}} = 8 \pi^2 R_A^2 = 18  \pi^2 (w+1)^2 t^2 \,,
\label{SQ}
\end{equation}
which diverges asymptotically, as $t \rightarrow \infty$.

\begin{figure}
\begin{center}
\center
\includegraphics[width=80mm]{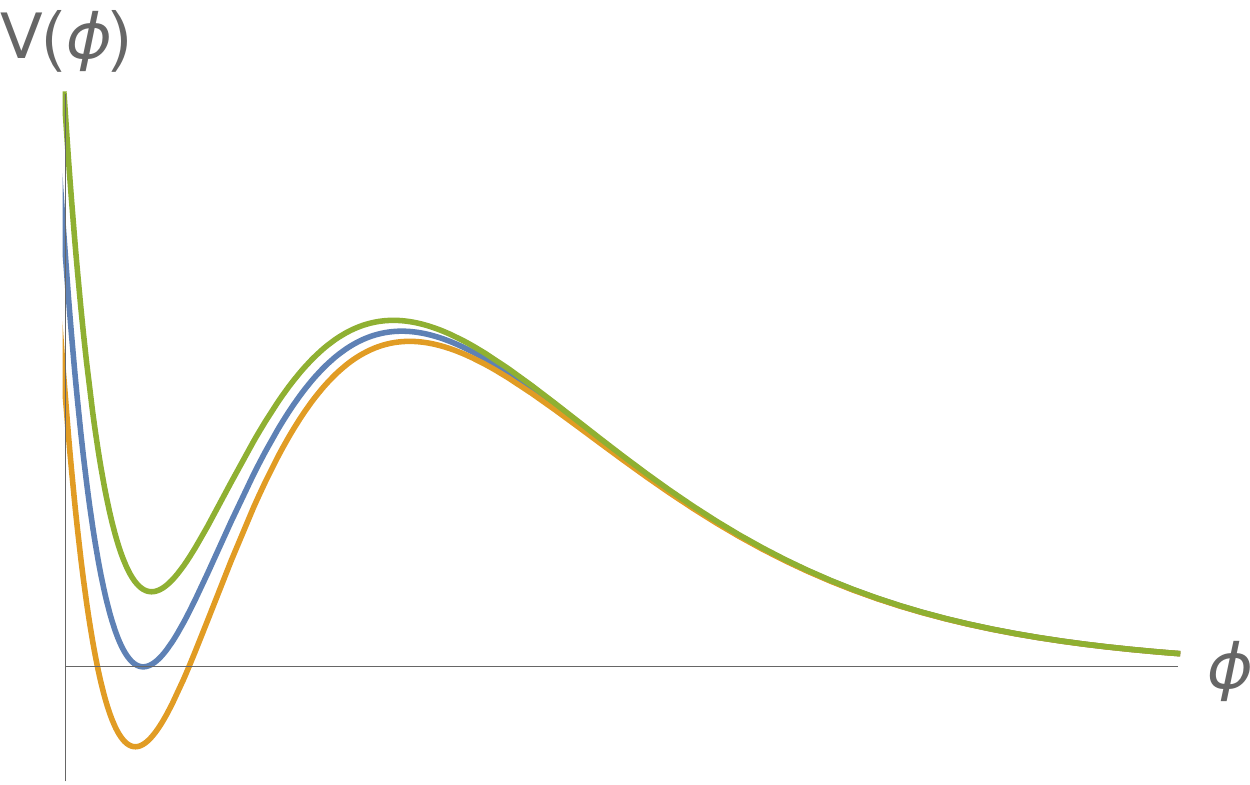}
\caption{Potentials with (a) a metastable dS minimum (green), (b) a Minkowski minimum (blue), and (c) an AdS minimum (yellow). CdL tunneling rates from the asymptotic region to any of these minima vanish as the field rolls to infinity.}
\label{potplot}
\end{center}
\end{figure}

By approximating Q-space as a metastable de Sitter minimum in the asymptotic region of a scalar field space, we may estimate its lifetime via the calculation of Coleman and de Luccia (CdL) \cite{Coleman:1977py}. In particular, let us suppose that the landscape contains another vacuum at finite $\phi$, which may have (a) $\Lambda > 0$, (b) $\Lambda = 0$, or (c) $\Lambda < 0$, as shown in Figure \ref{potplot}. For each of these possibilities, we claim that the CdL tunneling rate vanishes. 

This is easiest to see in the cases with $\Lambda > 0$ or $\Lambda = 0$. Up-tunneling from an initial vacuum to a final vacuum with larger cosmological constant ($\Lambda_i < \Lambda_f$) is exponentially suppressed by the difference in entropy between them,
\begin{equation}
\Gamma \sim \exp( S_f - S_i )\,.
\end{equation}
Here, the initial entropy $S_i$ diverges quadratically in time, by \eqref{SQ}. This means that the decay rate vanishes exponentially with $t^2$, and probability of decay in finite time is close to zero.

Similarly, Coleman and de Luccia showed that the rate of decay from a vacuum with $\Lambda_i > 0$ to one with $\Lambda_f = 0$ is suppressed (at least) by a factor of $\exp( - S_i )$. By the same reasoning as in the case of up-tunneling to de Sitter, the probability of decay in finite time is close to zero.

This leaves the possibility of a decay to an AdS minimum, with $\Lambda_f < 0$. At sufficiently late times, when $\Lambda_i = 18  \pi^2 (w+1)^2 t^2 \ll \Lambda_f$, 
such a tunneling is possible only if $3 T^2 < 4 \Lambda_f$, where $T$ is the tension of the domain wall between the two vacua.
This tension may be estimated as
\begin{equation}
T \sim |\Delta \phi| \sqrt{V_1}\,,
\end{equation}
where $V_1 > \Lambda_f$ is the height of the potential barrier between the two vacua. Thus, a nonzero tunneling rate requires 
\begin{equation}
|\Delta \phi|^2 < \frac{4 \Lambda_f}{3 V_1} < \frac{4}{3}\,.
\end{equation}
However, since we are (by assumption) far out in the asymptotic regime of scalar field space, we necessarily have $|\Delta \phi| \gg 1$. This implies that decay to an AdS minimum is similarly ruled out, and quintessence in asymptotic regimes of scalar field space is indeed stable, at least with respect to CdL tunneling.

It has long been argued that de Sitter vacua in quantum gravity can be at best metastable, since in stable de Sitter with a sufficiently small cosmological constant,\footnote{Bousso and Banks \cite{Banks:2007ei, Bousso:2011aa} have proposed $S_{\text{BB}} \approx 10^{25}$ as the minimum entropy needed to support a self-aware observer (also known as a ``Boltzmann brain''), which suggests that such observers will fluctuate into existence in de Sitter space with cosmological constant $\Lambda < 24 \pi^2/S_{\text{BB}} \approx 10^{-27}$. For our purposes, the precise number is irrelevant: the important thing is that it is some exponentially small, nonzero number.} the vast majority of observers will see a universe without an arrow of time, in blatant contradiction with our own observations \cite{Dyson:2002pf, Bousso:2011aa}. It is plausible that this argument also rules out stable Q-space, since at late times the effective cosmological constant $V$ will be arbitrarily close to zero, though further analysis of what constitutes an ``observer'' in Q-space is needed to make this precise.


\section{Conclusions}\label{sec:CONC}

In this paper, we have provided several lines of evidence for a bound of the form $\lH \geq \lm$ on the exponential coefficients of the Hubble scale and the mass scale of the lightest tower of particles, respectively, at late times in asymptotic regimes of scalar field space. We have further argued that the exponential coefficient of the Hubble scale is bounded above as $\lH \leq \sqrt{(d-1)/(d-2)}$. This complements previous works of the author, which have argued for bounds of the form $\lH \geq 1/\sqrt{d-2}$ and $\lm \geq 1/\sqrt{d-2}$. Together, these bounds offer a consistent picture for the late-time cosmology of asymptotic regimes of scalar field space in string theory: the Hubble scale $H$ remains at or below the Kaluza-Klein scale $\mKK$ and the string scale $\Mstring$, the universe experiences decelerated expansion $\ddot a \leq 0$, and the towers predicted by the Distance Conjecture are not parametrically lighter than the Hubble scale.

Our work raises a number of questions for future study. First and foremost, though the evidence we have provided for our proposed bounds is suggestive, it is not airtight. Either a proof or a counterexample to these bounds would be most welcome.

Another important question concerns the nature of the towers satisfying the Distance Conjecture.
In all the examples we have studied in this paper, the lightest tower of modes in any infinite-distance limit is either a tower of string oscillator modes or a tower of Kaluza-Klein modes, in accordance with the Emergent String Conjecture. However, while the evidence for the Emergent String Conjecture is quite formidable in the case of massless moduli spaces, it is not quite so clear that the Emergent String Conjecture applies to scalar fields with potentials. Indeed, even when the Emergent String Conjecture holds, it is possible in principle that lighter towers could exist with $\lm > \lKK, \lstring$. It would be worthwhile to find either an example of such a tower or else a stronger argument against such a possibility.

We have sketched one promising argument for the bound $\ddot a \leq 0$ in the previous section: stable Q-space may be ruled out for the same reason stable as de Sitter space in that both conflict with the observed arrow of time. The main obstacle facing this argument is the question of what exactly constitutes an ``observer'': it is known that arbitrarily large entropy fluctuations persist indefinitely in stable Q-space \cite{Bousso:2004tv}, but it is less clear that these fluctuations will spawn problematic self-aware observers. Further research on this question is needed to put this argument on firmer footing.

We have focused on the cosmological consequences for a universe with $\lH > \lKK$, but we have seen examples with $\lH = \lKK$. In this case, the late-time cosmology depends on subleading effects, which determine the relative size of the Hubble scale and the Kaluza-Klein scale. Further analysis of these subleading effects is therefore necessary to understand these borderline cases.

Finally, it must be emphasized that the bounds we have derived in various Kaluza-Klein/string compactifications apply to the asymptotic regime of scalar field space, where perturbative control allows for greater rigor. Since our own vacuum almost certainly lies in the interior of moduli space, it is not immediately clear how to apply the results of this paper to present-day cosmology. Extending our results beyond asymptotic regions to the interior of scalar field space is a challenging but promising step to bridging the gap between string theory and observation.


\section*{Acknowledgements}

We thank Ivano Basile, Raphael Bousso, Muldrow Etheredge, Ben Heidenreich, Sami Kaya, Miguel Montero, Yue Qiu, Matthew Reece, and Irene Valenzuela for useful discussions. This work was supported in part by the Berkeley Center for Theoretical Physics; by the Department of Energy, Office of Science, Office of High Energy Physics under QuantISED Award DE-SC0019380 and under contract DE-AC02-05CH11231; and by the National Science Foundation under Award Number 2112880.

\appendix

\section{Cosmology of Scalar Fields in Exponential Potentials}\label{App}

In this appendix, we review the cosmology of scalar fields in exponential potentials, generalizing the familiar analysis of quintessence in $d=4$ dimensions (see e.g. \cite{Copeland:1997et, Tsujikawa:2013fta}) to general $d$. In this context, we prove the bound $\lH \geq \sqrt{(d-1)/(d-2)}$.

We begin from the $d$-dimensional metric
\begin{equation}
ds^2 = -dt^2 + a(t)^2 d\vec{x}^2.
\end{equation}
and the action
\begin{equation}
S= \int d^d x \sqrt{-g} \left[ \frac{1}{2 \kappa_d^2}  \mathcal{R} - \frac{1}{2} (\nabla \phi)^2 -  V(\phi)  \right].
\end{equation}
Note that $\phi$ is canonically normalized, so we omit the $\hat\cdot$ throughout this appendix.
We further take $\phi$ to be homogenous, so its spatial derivatives vanish. The equation of motion for the scalar field is thus
\begin{equation}
\ddot {\phi} + (d-1) H \dot { \phi } = - V'(\phi)\,,
\end{equation}
where $H = \dot a/a$.
The stress-energy tensor is that of a perfect fluid of density $\rho$, pressure $p$, with
\begin{equation}
\rho = \frac{1}{2} \dot {\phi}^2 + V(\phi)\,,~~~ p  = \frac{1}{2} \dot {\phi}^2 - V(\phi)\,.
\label{rhogen}
\end{equation}
The Friedmann equations take the form:
\begin{equation}
H^2 = \frac{2 \rho \kappa_d^2}{(d-1)(d-2)} \,,~~~ \dot H = - \frac{p \kappa_d^2}{(d-2) } - \frac{(d-1) H^2}{2 }\, .
\label{Fgen}
\end{equation}

Next, we assume that $V$ takes an exponential form,
\begin{equation}
V(\phi) = V_0 \exp( - c \kappa_d \phi  )\,.
\label{Vexpform}
\end{equation}
Extending the analysis of \cite{Copeland:1997et} (see also \cite{Tsujikawa:2013fta}) to $d$ dimensions, we define dimensionless parameters
\begin{equation}
x = \frac{ \kappa_d \dot \phi }{H \sqrt{(d-1)(d-2)}} \,,~~~~y = \frac{\kappa_d \sqrt{2 V} }{ H \sqrt{(d-1)(d-2)} }\,.
\end{equation}
Assuming henceforth that the only source of stress-energy is the rolling scalar field, these satisfy $x^2+y^2=1$, by \eqref{rhogen}, \eqref{Fgen}.
Defining the dimensionless time parameter $N$ by $dN  =  H dt$, the equation of motion for $\phi$ yields
\begin{align}
\frac{dx}{dN} &= \frac{c y^2 \sqrt{ (d-1)(d-2) }}{2}  - (d-1) x y^2 \,,\\
\frac{dy}{dN} &= - \frac{c x y\sqrt{ (d-1)(d-2) }}{2}  + (d-1) y (1-y^2)  \,.
\end{align}
This dynamical system features two relevant fixed points:\footnote{$(x, y) = (0,0)$ is also a fixed point, but it does not satisfy the constraint $x^2+y^2=1$. $(x,y) = (-1,0)$ is also a fixed point, but the choice of sign in the potential in \eqref{Vexpform} implies $x > 0$ at late times.}
\begin{equation}
(x_1, y_1) = ( 1, 0) \,,~~~(x_2, y_2) = \left(\frac{c}{2} \sqrt{\frac{d-2}{d-1} }, \sqrt{ 1 - \frac{c^2(d-2)}{4(d-1)}} \right)\,.
\end{equation}
Note that the second fixed point exists only for $c \leq 2 \sqrt{(d-1)/(d-2)}$, and it coalesces with the first fixed point for $c  = 2 \sqrt{(d-1)/(d-2)}$. Whenever it exists, it is a late-time attractor, which leads at late times to an equation of state parameter
\begin{equation}
w \equiv p/\rho = -1 + \frac{1}{2} \frac{d-2}{d-1}  c^2\,.
\label{wgen}
\end{equation}
By \eqref{wgen} and \eqref{Fgen}, the Hubble parameter scales simply with $\phi$ as
\begin{equation}
H \sim \sqrt{V} \sim \exp( - c \kappa_d \phi/2 )\,,
\end{equation}
so $\lH = c/2$.

Meanwhile, for $c  > 2 \sqrt{(d-1)/(d-2)}$, the first fixed point $(x_1, y_1) = (1,0)$ is a late-time attractor, which leads at late times to an equation of state parameter
$
w \equiv p/\rho = 1$.

To find $\lH$ at late times, where $(x, y) \rightarrow (1, 0)$, we set $V=0$ in \eqref{wgen} and \eqref{Fgen}, yielding
\begin{equation}
\dot H = - \frac{\dot \phi^2 \kappa_d^2}{d-2} = -  \dot \phi  H \kappa_d \sqrt{ \frac{d-1}{d-2}}\,. 
\end{equation}
This implies
\begin{equation}
\frac{d}{dt} (\log H) = - \kappa_d \sqrt{ \frac{d-1}{d-2}} \frac{d}{dt} \phi\,,
\end{equation}
so
\begin{equation}
H \sim \exp\left(  -  \kappa_d \sqrt{ \frac{d-1}{d-2}}  \phi \right)\,,
\end{equation}
hence $\lH = \sqrt{(d-1)/(d-2)}$.

Putting this all together, we find that $\lH = c/2$ for $c \leq 2 \sqrt{(d-1)/(d-2)}$, and $\lH = \sqrt{(d-1)/(d-2)}$ for $c > 2 \sqrt{(d-1)/(d-2)}$. This can be summarized succinctly as
\begin{equation}
\lambda_H = \min \left( \frac{c}{2}, \sqrt{\frac{d-1}{d-2}} \right)\,.
\end{equation}

Meanwhile, we have
\begin{equation}
w = \min\left( -1 + \frac{1}{2} \frac{d-2}{d-1}  c^2 , 1 \right)\,.
\end{equation}
This implies $|w| \leq 1$ at asymptotically late times, which is equivalent to the dominant energy condition. Indeed, the dominant energy condition follows immediately from \eqref{rhogen}.

If we further assume the bound $c \geq 2/\sqrt{d-2}$, as advocated above, then 
\begin{equation}
 w \geq - \frac{d-3}{d-1}\,, 
\end{equation}
which is equivalent to the strong energy condition.

In theories with exponential potentials involving multiple (canonically normalized) scalar fields, our analysis here suggests (and simple numerical studies confirm) that the late-time dynamics similarly depends on whether the gradient of the potential in asymptotic regions satisfies the inequality $|\nabla V| / V \leq 2 \sqrt{(d-1)/(d-2)}$. If this inequality is satisfied, then as in the single field case studied here, the dynamics of the rolling scalar field will be subject to the pull of the gradient of the potential, and at late times the field will approach a gradient descent trajectory with $\lH = |\nabla V|/(2 V)$. If, on the other hand, the potential satisfies the inequality $|V'| / V \geq 2 \sqrt{(d-1)/(d-2)}$, where $V' <0$ is the directional derivative of the potential along some direction in scalar field space, then at late times the potential may be neglected, and the field may roll to infinity along the direction of interest with $\lH = \sqrt{(d-1)/(d-2)}$, $w = 1$.


\bibliographystyle{utphys}
\bibliography{ref}

\end{document}